%% file: main.tex
\documentclass[sigconf, nonacm]{acmart}
\usepackage{caption}

\AtBeginDocument{%
  }

\usepackage{graphicx}
\usepackage{multirow}
\usepackage[table]{xcolor}
\usepackage{array}
\usepackage{hhline}
\usepackage{makecell}
\usepackage{subcaption}
\usepackage{booktabs}
\usepackage{pifont}
\usepackage{amsmath}
\makeatletter
\@ifpackageloaded{newtxmath}{}{\usepackage{amssymb}}
\makeatother
\usepackage{tikz}
\usepackage{xspace}
\providecommand{\checkmark}{\ding{51}}

\begin{document}

\title{HoliBench: A Cross-Platform Benchmarking and Deployment Toolkit for Foundation Models in CPS-IoT Applications}

\author{Inesh Chakrabarti}
\authornote{Both authors contributed equally to this research.}
\email{inesh33@ucla.edu}
\affiliation{%
  \institution{University of California, Los Angeles}
  \city{Los Angeles}
  \state{California}
  \country{USA}
}

\author{Zejun Xiong}
\authornotemark[1]
\orcid{0009-0008-9152-7268}
\email{zejunxiong@ucla.edu}
\affiliation{%
  \institution{University of California, Los Angeles}
  \city{Los Angeles}
  \state{California}
  \country{USA}
}

\author{Pragya Sharma}
\email{pragyasharma@ucla.edu}
\affiliation{%
  \institution{University of California, Los Angeles}
  \city{Los Angeles}
  \state{California}
  \country{USA}
}

\author{Mani Srivastava}
\email{mbs@ucla.edu}
\affiliation{%
  \institution{University of California, Los Angeles}
  \city{Los Angeles}
  \state{California}
  \country{USA}
}

\renewcommand{\shortauthors}{Chakrabarti et al.}

\begin{abstract}
Foundation models, including large language models, vision-language models, and time-series foundation models, are increasingly deployed on embedded and edge platforms for CPS and IoT applications, where energy, latency, and memory are as critical as task accuracy. Existing benchmarking tools evaluate model capability in isolation, reporting accuracy under the assumption of sufficient compute, while hardware profiling tools remain platform-specific and mutually incompatible. As a result, users lack a unified workflow for comparing configurations and making deployment decisions across heterogeneous devices.

We present HoliBench, a modular benchmarking and deployment toolkit that jointly characterizes accuracy, latency, and energy across platforms from single-board computers to GPU servers. Its platform abstraction layer calibrates cross-device measurement, and the toolkit supports multiple model modalities, inference engines, concurrencies, and existing evaluation harnesses. An interactive interface exposes constraint-aware configuration selection over a design space that is profiled once and reused across studies.

Using HoliBench, we characterize 20 models across 7 device types, 3 quantization levels, 8 inference backends, and over 30 tasks, surfacing tradeoffs that existing tools miss: quantization reduces latency only on hardware with low-precision support, accuracy gains show diminishing returns relative to energy, and for autoregressive workloads, average inference power is approximately constant across output lengths. We further find that single-model profiles compose under sequential co-resident execution. In a multi-model CPS deployment, standalone profiles predict combined-pipeline latency and power within 1.2\% and 2.5\%, enabling deployment exploration without exhaustive profiling of every pipeline configuration. HoliBench surfaces feasible configurations that accuracy-only evaluation cannot identify. We release HoliBench\footnote{Additional results, demos and interactive UI: \url{https://github.com/beesfleas/HoliBench}} as open-source infrastructure for deployment-aware evaluation of foundation models.
\end{abstract}

\ccsdesc[500]{Computer systems organization~Embedded systems}
\ccsdesc[300]{Computing methodologies~Neural networks}
\ccsdesc[100]{Software and its engineering~Software performance}

\keywords{foundation models, benchmarking, edge computing, CPS, IoT, energy efficiency, quantization, deployment-aware evaluation}

\maketitle

\input{sections/introduction}
\input{sections/related_work}
\input{sections/system_architecture}
\input{sections/technical_impl}
\input{sections/exp_setup}
\input{sections/evaluation}
\input{sections/case_study}
\input{sections/conclusion}

\begin{acks}
{\sloppy
This research was funded by DEVCOM ARL under award \#W911NF1720196, NIH under award \#1P41EB028242, and NSF under award CNS \#2325956.
The authors used generative AI tools, including Codex/\allowbreak ChatGPT, to assist with software development tasks such as documentation, unit test generation, boilerplate implementation, debugging, and parts of the codebase implementation. The authors also used generative AI tools during manuscript preparation for brainstorming, outlining, and editing assistance. The authors conceived the architecture, designed the study, conducted the experiments, and verified all AI-assisted code and text. The authors take full responsibility for the final artifact and manuscript.\par}
\end{acks}

\bibliographystyle{ACM-Reference-Format}
\bibliography{ref}

\end{document}

%% file: sections/introduction.tex
\section{Introduction}
\label{sec:intro}

Embedded and cyber-physical systems have historically relied on task-specific deep neural networks, with each application requiring its own trained model, optimization pipeline, and deployment workflow. Foundation models (FMs) offer a fundamentally different tradeoff where a single pretrained model, adapted through lightweight mechanisms such as prompting or finetuning~\cite{brown2020gpt3, hu2022lora}, can serve tasks that previously demanded separate specialized DNNs for vision, language, and sensor data processing~\cite{bommasani2021foundation}. Recent work has validated this approach on resource-constrained hardware. Multimodal FMs can operate as on-device firmware on smartphones~\cite{yuan2024m4} or enable open-set learning on embedded platforms~\cite{yang2024edgefm}, and quantization methods such as AWQ~\cite{lin2024awq} and GPTQ~\cite{frantar2023gptq} bring 1B to 7B parameter models within the memory and compute envelopes of edge accelerators and single-board computers (SBCs).

The question facing users today is no longer whether foundation models can run at the edge, but which model, runtime, quantization configuration, and target device should be selected for a deployment under latency, energy, memory, and accuracy constraints. There is currently no unified tooling that allows users to answer this question end-to-end. Accuracy-focused benchmarking suites, including EleutherAI's LM Evaluation Harness~\cite{biderman2024lmeval} and HELM~\cite{liang2023helm}, evaluate model capability across standardized tasks but are agnostic to the hardware that executes the model. They report perplexity and task accuracy assuming sufficient compute, which reveals nothing about whether a given configuration will meet latency or energy constraints on a target device. MLPerf Inference~\cite{reddi2020mlperf} provides rigorous throughput and latency measurements and has recently incorporated FM workloads, but its suite is fixed, its edge category still assumes relatively capable hardware, and it does not collect energy data. Characterization efforts like MELT~\cite{laskaridis2024melt} and the recent study by Abstreiter et al.~\cite{abstreiter2026ondevice} profile LLM inference on specific edge platforms, providing useful but narrow empirical snapshots. Each toolkit uses its own measurement setup and covers a limited slice of the design space, producing results that are not directly comparable.

This gap is particularly consequential for embedded AI and on-device foundation model deployment, where latency and energy are not secondary to accuracy but first-class deployment constraints. The pressure is acute in CPS and IoT settings, where a vision-language model (VLM) guiding robotic navigation must hold inference within a tight latency window under a sustained power budget, and a time-series foundation model (TSFM) processing sensor streams must produce predictions faster than the sampling rate within a few gigabytes of memory. Users must jointly reason about accuracy, latency, power, and memory across candidate configurations, yet in practice this requires stitching together incompatible tools. A typical workflow runs an accuracy harness such as lm-eval-harness on one machine, separate power-sampling scripts on each target device, and merges the resulting logs by hand.

Several practical challenges make cross-platform evaluation of foundation models difficult. First, the foundation model landscape is not a single workload class. Autoregressive LLMs, vision-language models, and time-series transformers place fundamentally different demands on the hardware they run on. LLM decode is memory-bandwidth-bound with a sequentially growing KV-cache, while VLMs front-load a compute-intensive vision encoding stage before autoregressive generation begins. TSFMs bypass autoregressive generation entirely, operating over fixed-length numerical windows in a single forward pass. These structural differences mean that a single aggregate latency or energy measurement per model obscures the bottleneck that actually governs deployment feasibility. Profiling that does not account for these differences will produce comparisons that do not generalize across modalities.

Second, heterogeneous hardware platforms expose fundamentally different APIs for the measurements that matter most in constrained deployment, including power consumption, memory pressure, and processor utilization. The interfaces, sampling granularity, and measurement semantics differ across ARM-based SBCs, GPU-capable devices, and unified memory-enabled platforms. Without a unified abstraction layer that accounts for these differences, raw numbers from different platform APIs cannot be compared directly. They must be normalized, calibrated, and temporally aligned with inference events to produce commensurable results.

Third, even given accurate cross-platform measurements across diverse FM workloads, the resulting design space is large and difficult to navigate manually. A user evaluating a modest set of 10 models across 4 quantization levels on 5 devices already faces 200 configurations before considering task-specific accuracy requirements and operational constraints on power, latency, and memory. Additionally, these constraints interact, e.g., a larger model may deliver higher accuracy but exceed the latency budget on a given device. Practical decision support, rather than manual trial and error, is needed to navigate these intertwined tradeoffs.

More fundamentally, they raise a systems question. Can measurements collected from isolated foundation-model executions predict the behavior of larger multi-model deployments, or must every candidate deployment be profiled independently?

These challenges motivate the need for a reusable, end-to-end workflow that supports consistent measurement, comparison, and configuration selection across heterogeneous platforms.

We present HoliBench, a cross-platform \textbf{holi}stic \textbf{bench}marking and deployment toolkit for foundation models spanning the device-edge-cloud continuum. HoliBench is designed as deployment-time infrastructure for users evaluating model feasibility under real-world constraints, supporting workflows that combine offline profiling, cross-platform comparison, and constraint-aware configuration selection. Beyond providing unified measurement infrastructure, HoliBench enables characterization of deployment behaviors that remain invisible to capability-centric evaluation alone. This paper makes the following contributions.

\begin{enumerate}
\item We present HoliBench, an open-source benchmarking and deployment toolkit for foundation models spanning LLMs, VLMs, and TSFMs across heterogeneous hardware from SBCs to GPU servers. HoliBench supports multiple inference engines, quantization backends, and evaluation harnesses through a modular and extensible architecture.

\item We show that standalone foundation-model profiles compose under sequential co-resident execution. In a multi-model CPS deployment, the profiles predict combined-pipeline latency and power with errors of 1.2\% and 2.5\%, enabling deployment exploration without exhaustively profiling every candidate pipeline configuration.

\item HoliBench provides a platform abstraction layer that enables unified power, energy, memory, and utilization profiling across heterogeneous hardware through synchronized measurement, temporal alignment, and consistent metrics.

\item We characterize foundation-model deployment across 20 models, 7 devices, and 3 quantization levels, and show that deployment behavior often diverges from common expectations: quantization benefits depend strongly on runtime and hardware support and can reverse into latency penalties, average inference power is largely determined by the target device rather than model size, and accuracy gains exhibit diminishing returns relative to energy cost.

\item HoliBench enables constraint-aware configuration selection through an interactive interface that identifies feasible model-device-quantization configurations under joint accuracy, latency, energy, and memory constraints from a reusable profiled design space.
\end{enumerate}

%% file: sections/related_work.tex
\section{Related Work}

\subsection{Foundation Model Deployment and Characterization}

Recent work has established that foundation models can run effectively on resource-constrained hardware (Section~\ref{sec:intro}). Post-training quantization methods including AWQ and GPTQ, together with weight-staging techniques such as PowerInfer-2~\cite{xue2024powerinfer2}, have brought 1B to 7B parameter models within the memory and compute envelopes of edge accelerators and smartphones. As deployment has become feasible, several studies have begun characterizing FM inference on specific devices. MELT~\cite{laskaridis2024melt} measured latency, memory, and energy for LLMs across mobile and edge hardware but relied on a single inference backend (\texttt{llama.cpp}) and did not extend to VLMs or TSFMs. Abstreiter et al.~\cite{abstreiter2026ondevice} produced a detailed tradeoff analysis on Raspberry Pi~5 and Jetson Orin Nano, covering power and throughput under various quantization configurations. Lu et al.~\cite{lu2025slm} focused on quantization and pruning strategies for sub-3B language models at the edge. While each study contributes useful data, they construct independent measurement infrastructure, cover disjoint slices of the model-device space, and define metrics and baselines differently, making cross-study comparison difficult.

\subsection{FM Capability Benchmarks}

The model evaluation community has developed frameworks for assessing FM capability, including HELM~\cite{liang2023helm}, Big-Bench~\cite{srivastava2022bigbench}, Chatbot Arena~\cite{chiang2024chatbot}, the Hugging Face Open LLM Leaderboard~\cite{beeching2024leaderboard}, and EleutherAI's LM Evaluation Harness~\cite{biderman2024lmeval}. These frameworks operate under an implicit assumption that compute is abundant. They report accuracy, perplexity, and calibration but do not measure what those scores cost in inference latency or energy on a given device. For deployment on embedded and edge platforms, where power budgets are typically low, capability metrics alone do not determine whether a model configuration is viable.

\subsection{Hardware-Aware Profiling Infrastructure}

MLPerf Inference~\cite{reddi2020mlperf} is the established benchmark for datacenter and edge inference throughput and latency. Recent rounds include FM workloads, but the suite is fixed by committee, the edge division assumes relatively capable hardware, and energy metrics fall outside its scope. MLPerf Tiny~\cite{banbury2021mlperftiny} extends coverage to microcontroller-class devices with traditional DNN workloads rather than foundation models. Roofline-style analytical models~\cite{zhou2023pagoda} and micro-NPU characterization efforts~\cite{reddi2023micronpu} contribute performance reasoning tools for specific hardware classes but do not constitute cross-platform measurement infrastructure. Across this body of work, no single system provides unified measurement of latency and energy for FM inference spanning the device-cloud spectrum.
Sharma et al.~\cite{sharma2025icccn} further show that deployment placement decisions depend on workload-specific latency and resource constraints rather than model capability alone.

\medskip
\noindent HoliBench addresses the intersection of these three areas. It provides cross-platform measurement infrastructure with a hardware abstraction layer that standardizes energy, performance, and latency profiling across heterogeneous devices, from embedded devices to GPU servers. Its support for LLMs, VLMs, and TSFMs along with various inference engines covers modalities absent from all prior profiling and characterization work. Its configuration-selection layer exposes the profiled design space to constraint-aware exploration through a pluggable solver interface, a capability absent from prior profiling and benchmarking frameworks.

%% file: sections/system_architecture.tex
\begin{figure}[t]
\centering
\includegraphics[width=\columnwidth, trim=0 1650 0 0, clip]{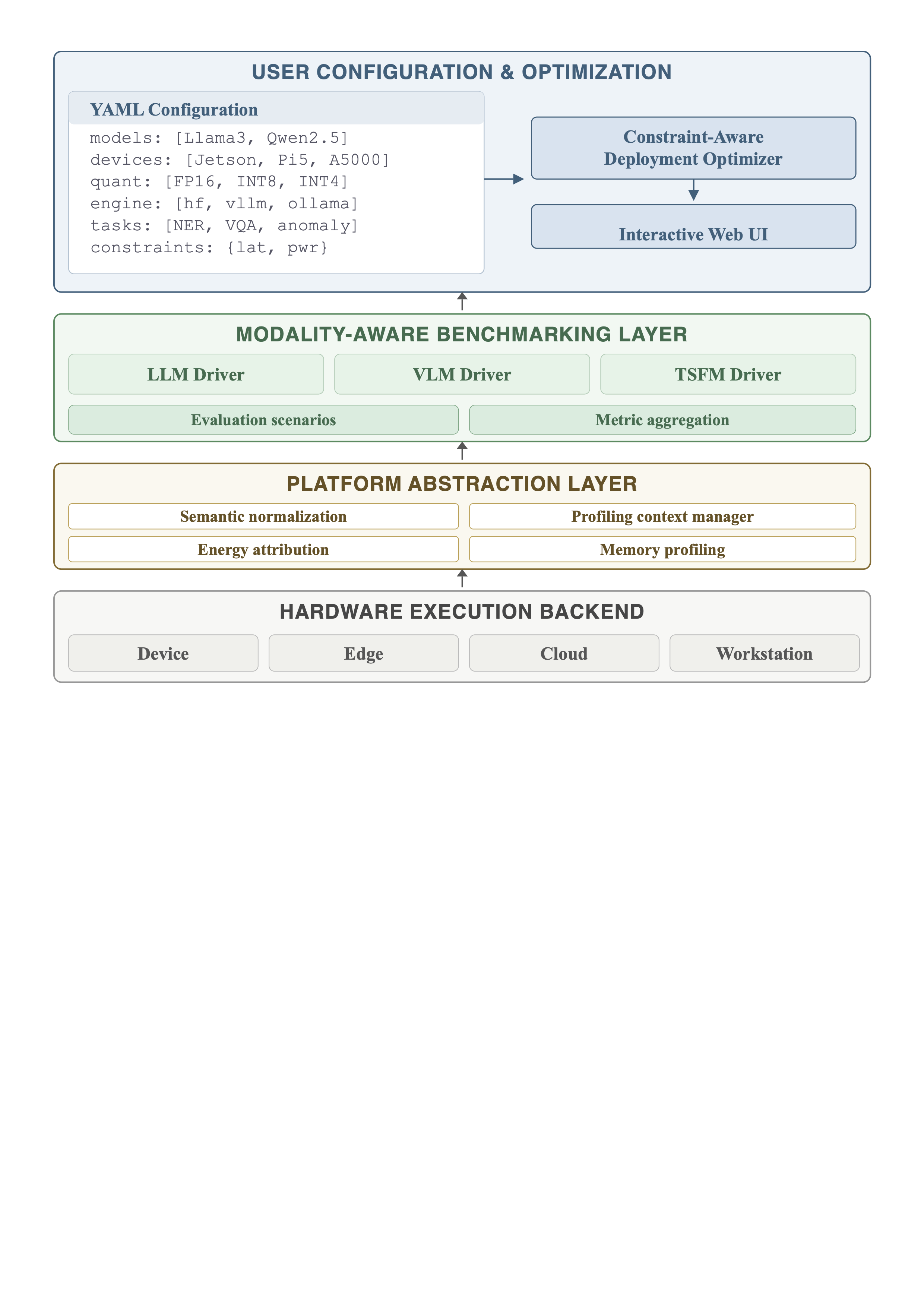}
\caption{HoliBench system architecture. User-specified configurations feed into the constraint-aware configuration selector. Note that configuration selection is the broader concept, with deployment optimization as one realization implemented by HoliBench's solver. Modality-aware drivers profile FM workloads, and the platform abstraction layer normalizes heterogeneous telemetry across four hardware deployment tiers.}
\label{fig:arch}
\end{figure}

\section{System Architecture}
\label{sec:architecture}

The HoliBench architecture is structured as a modular, four-layer hardware-in-the-loop stack designed to address the challenges of measurement heterogeneity, modality-specific execution patterns, and combinatorial deployment optimization. The user specifies models, quantization levels, tasks, target devices, and inference engine through a YAML-driven configuration interface. New models, devices, and evaluation tasks are integrated by implementing fixed interface contracts at the relevant layer, without modifying the orchestration logic, making HoliBench extensible by design.

\subsection{Platform Abstraction Layer (PAL)}
\label{sec:pal}

The primary challenge in cross-platform characterization is the lack of semantic commensurability between hardware measurement interfaces. HoliBench abstracts platform-specific telemetry into a unified Platform Abstraction Layer that ensures physical units are derived through consistent integration methods.

\textbf{Temporal Alignment.}
To prevent measurement drift and ensure precise attribution, HoliBench employs a Profiling Context Manager. This component enforces synchronous bracketing of the inference workload across all active telemetry background threads, ensuring that energy and utilization traces correspond strictly to the computational window and eliminating the inclusion of extraneous idle-state activity.

\textbf{Semantic Normalization.}
The framework distinguishes between raw signal acquisition and derived metric computation. For interfaces providing instantaneous power $P$, such as NVIDIA NVML or the Raspberry Pi PMIC, total energy $E$ is computed by integrating instantaneous power over the inference window:
\begin{equation}
E = \int_{t_{\text{start}}}^{t_{\text{end}}} P(t)\, dt
\label{eq:energy_instantaneous}
\end{equation}

Conversely, for architectural interfaces providing cumulative energy counters (e.g., Intel RAPL), energy is calculated via the differential:
\begin{equation}
E = C(t_{\text{end}}) - C(t_{\text{start}})
\label{eq:energy_cummulative}
\end{equation}

\textbf{Energy Attribution.} To isolate the marginal cost of inference, the framework implements a differential power analysis phase. A baseline idle power $P_{\text{idle}}$ is characterized post-model-loading but pre-inference. HoliBench then reports both Gross Energy (total system draw during the inference window) and Net Inference Energy:
\begin{equation}
E_{\text{net}} = E_{\text{gross}} - \int_{t_{\text{start}}}^{t_{\text{end}}} P_{\text{idle}}\, dt
\label{eq:energy_net}
\end{equation}

This separation enables fair efficiency comparisons across devices with vastly different static power envelopes. Gross energy captures the operational cost relevant to power budgeting while net inference energy isolates the marginal computational cost and is the appropriate basis for cross-device comparison.

\textbf{Memory Profiling.}
Host memory and device memory (VRAM on discrete GPUs, unified memory on SoC platforms) are tracked with peak and time-averaged utilization recorded per run. Peak memory is the binding feasibility constraint, determining whether a model at a given quantization level fits on the target device. Average memory characterizes sustained pressure during inference, which is relevant when concurrent workloads share a device.

\begin{figure}[t]
\centering
\includegraphics[width=1\columnwidth]{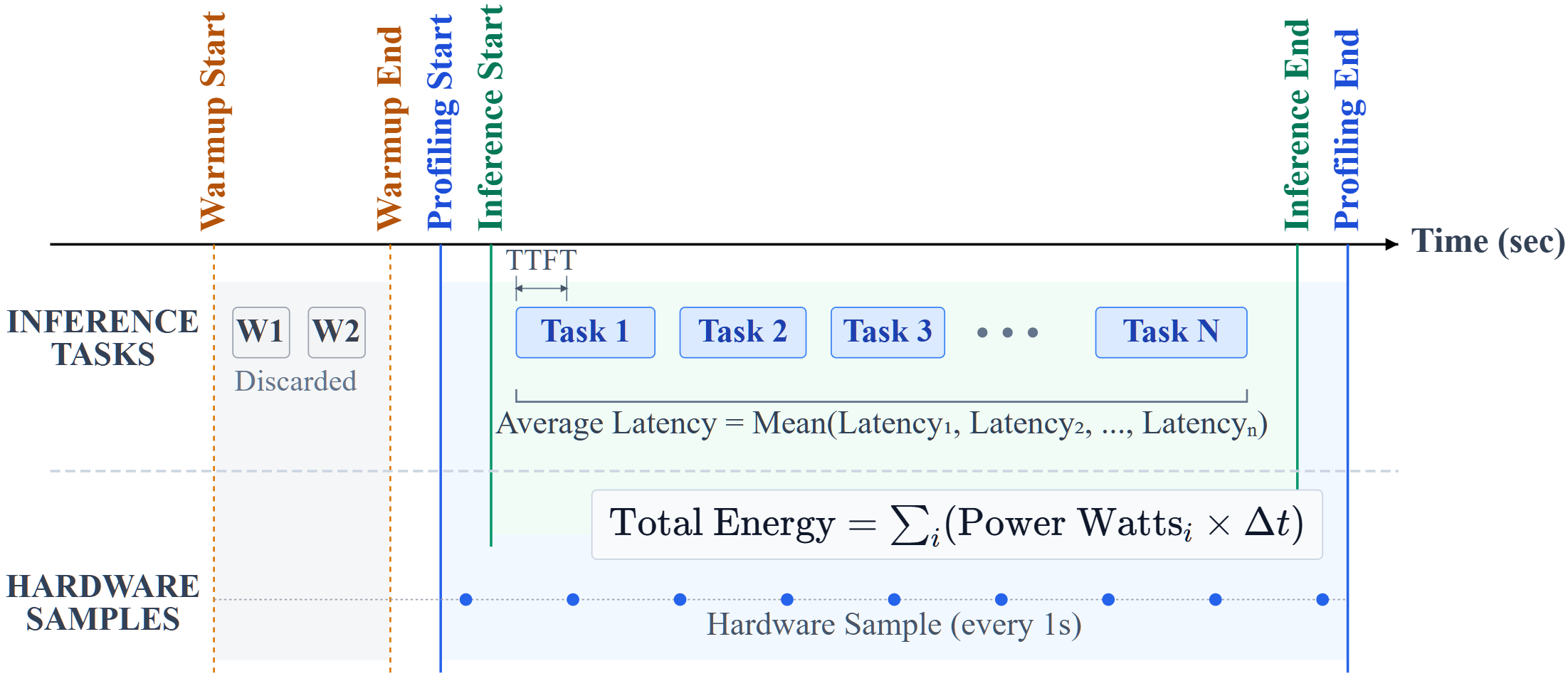}
\caption{Benchmarking timing pipeline including warmup, timed inference tasks (TTFT, latency), and concurrent hardware power sampling. (Note that the sampling period can range from 10ms-1s based on the model under test.)}
\vspace{-1.5em}
\label{fig:macro_timeline}
\end{figure}

\subsection{Inference Timing Model}
\label{sec:timing-model}
Every benchmark run proceeds through a warmup phase
that primes runtime caches followed by
a profiling window in which the context manager activates
all telemetry threads and dispatches $N$ timed inference
tasks (Figure~\ref{fig:macro_timeline}). Warmup results
are discarded and aggregate metrics are computed only
over the measured tasks.

\begin{figure}[t]
\centering
\includegraphics[width=1\columnwidth]{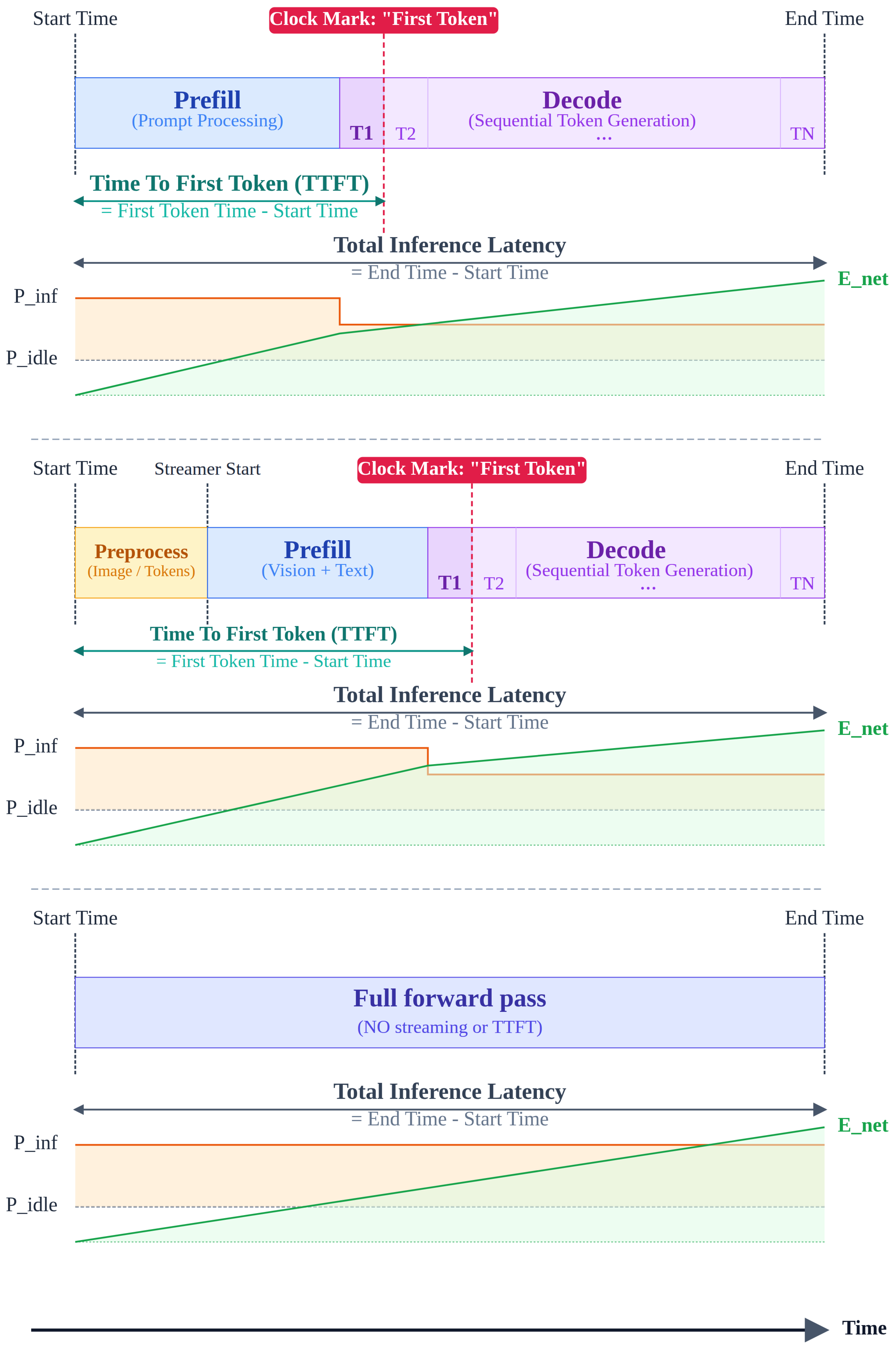}
\caption{Inference timing model for the three FM workload
classes. Each panel shows profiler brackets, power sampling,
$P_{\text{inf}}$, $P_{\text{idle}}$, and $E_{\text{net}}$.}
\vspace{-2em}
\label{fig:timing}
\end{figure}

Although FM inference shares a common macro-level temporal structure (Figure~\ref{fig:macro_timeline}), the per-task execution phases differ across modalities and determine how the PAL attributes latency and energy to each stage (Figure~\ref{fig:timing}).
For autoregressive LLMs, a compute-bound prefill pass processes
all input tokens, populates the KV-cache, and produces the first
output token $T_1$ at the TTFT clock mark.
Subsequent tokens $T_N$ are generated per decode
step, yielding total inference latency
$T = T_{\text{prefill}} + T_{\text{decode}}$.
VLMs exhibit a two-phase structure where a compute-bound vision
encoder processes the input image before autoregressive language
generation begins. HoliBench records the phase boundary to
decompose $T = T_{\text{vision}} + T_{\text{decode}}$.
TSFMs bypass autoregressive generation entirely, completing
inference in a single forward pass with no KV-cache and no
phase decomposition (Figure~\ref{fig:timing}).

In all three cases, the profiling context manager brackets the
inference window and samples power at regular intervals.
Net inference energy subtracts the idle baseline, characterized
post-model-load but pre-inference, from the gross power integral
(Eq.~\ref{eq:energy_net}).
This phase-level decomposition enables the latency interpolation
and time-weighted power estimation used in the multi-model case
study (Section~\ref{sec:eval-composability}).

\subsection{Modality-Aware Benchmarking}
\label{sec:modality}

HoliBench supports inference through the model's native forward pass as well as through 8 separate higher-level serving runtimes and engines including vLLM~\cite{kwon2023vllm}, SGLang \cite{zheng2024sglang}, HuggingFace, MLX \cite{mlx2023}, OpenVino \cite{openvino_toolkit}, and TensorRT \cite{tensorrt}, ollama \cite{ollama2024}, and ONNX\cite{onnxruntime}. Native forward-pass inference dispatches directly to the execution backend available on the target platform (e.g., CUDA on NVIDIA GPUs, CPU fallback on ARM), ensuring that measured latency and energy are attributable to the model-hardware interaction rather than to runtime-level scheduling or memory management policies. We also support batching and evaluating batched performance (decode vs prefill) with metrics like prefill throughput and decode throughput at differing contexts.

\textbf{Quantization Sensitivity.}
The infrastructure evaluates the cross-product of quantization precision (FP16, INT8, INT4) and hardware-specific acceleration support. While HoliBench's characterization uses bitsandbytes weight-only quantization, it also supports other popular methods allowing users to profile the same model under different quantization backends on the same device. Because quantization interacts with hardware architecture in ways that cannot be predicted from model compression ratios or device specifications alone, HoliBench profiles the actual quantized execution path on each target device rather than assuming uniform benefit from reduced bit-width.

\textbf{Evaluation.}
Each task is encapsulated as a modality-specific scenario that defines dataset loading, input formatting, and scoring against task-appropriate metrics. This separation ensures that profiling and metric aggregation remain shared across modalities while evaluation protocols adapt to each workload's output semantics.

\subsection{Constraint-Aware Configuration Selection}
\label{sec:optimizer}

\begin{table*}[!ht]
\centering
\small
\begin{tabular}{llllcc}
\toprule
\textbf{Platform} & \textbf{Tier} & \textbf{Processor} & \textbf{Memory} & \textbf{Profiler} & \textbf{Tensor} \\
 & & & & \textbf{Backend} & \textbf{Cores} \\
\midrule
Raspberry Pi 5~\cite{raspberrypi5} & Device & BCM2712 (ARM) & 8\,GB& PMIC & -- \\
Jetson Orin AGX~\cite{nvidia_jetson_orin_agx} & Edge & Ampere GPU + ARM & 64\,GB& jtop & $\checkmark$ \\
RTX 3070~\cite{nvidia_rtx3070} & Edge & Ampere GPU & 8\,GB& NVML & $\checkmark$ \\
RTX 5070 ~\cite{nvidia_rtx5070} & Cloud & Blackwell GPU & 12\,GB& NVML &  $\checkmark$  \\
RTX A5000~\cite{nvidia_rtx_a5000} & Cloud  & Ampere GPU  & 24\,GB  & NVML  & \checkmark \\
Mac Mini M2~\cite{apple_mac_mini_m2} & Workstation & Apple M2 Pro (unified) & 16\,GB& powermetrics & -- \\
Mac Mini M4~\cite{apple_m4} & Workstation & Apple M4 Pro (unified) & 24\,GB& powermetrics & -- \\
\bottomrule
\end{tabular}
\caption{Hardware testbed for HoliBench. The Memory
column reports GPU VRAM for discrete NVIDIA GPUs, unified CPU/GPU
memory for Apple Silicon and Jetson platforms, and system RAM for
the Raspberry Pi~5.}
\vspace{-1em}
\label{tab:platforms}
\end{table*}

The final layer of HoliBench supports constraint-aware selection of deployment configurations from the profiled design space. Given the profile table produced by the preceding layers, indexed by model, quantization level, inference backend, and device, together with user-specified constraints on latency, power, memory, and accuracy, and weights across the objectives, this layer returns a ranked set of feasible configurations.

The layer is structured as two stages. The first stage performs solver-agnostic feasibility pruning, eliminating configurations that violate hard constraints before any selection logic runs. These include hardware limits (e.g., a model whose memory footprint at a given quantization level exceeds the target device's VRAM) and modality requirements (e.g., an LLM cannot serve a visual question-answering task). Peak memory is treated strictly as a hard constraint and is never traded against other objectives, since a configuration that does not fit on the device cannot be deployed regardless of its other properties. The surviving configurations form the candidate set passed to the second stage.

The second stage is a pluggable solver. The solver interface accepts any implementation conforming to the contract \\ \texttt{solve(candidate\_set, constraints, weights)} $\rightarrow$ \\ \texttt{ranked\_configurations}, separating the calibrated measurement infrastructure (the profile table) from the selection policy applied to it. HoliBench ships with an Integer Linear Programming (ILP) solver as the representative implementation. The ILP defines a weighted utility $U_i$ over normalized metrics for each configuration $i \in \mathcal{C}$ and selects the configuration maximizing $U_i$, subject to the user-specified deployment constraints. Formally,
\begin{equation}
\text{Maximize } U_i = \sum_j w_j \hat{v}_{i,j} \quad\text{subject to:}
\end{equation}
\begin{equation*}
l_i \leq L_{\max}, \quad e_i \leq E_{\max}, \quad m_i \leq M_{\max},
\quad a_i \geq A_{\min}
\end{equation*}
where $\hat{v}_{i,j}$ is the normalized value of configuration $i$ on objective $j$, $w_j$ is the user-defined weight on that objective, and $l_i$, $e_i$, $m_i$, $a_i$ are the configuration's measured latency, energy, peak memory, and accuracy. Peak memory remains a hard constraint and is excluded from the utility score. The same utility ranks the remaining feasible configurations, producing the ordered output exposed to the user.

The same interface admits other selection strategies such as random search or greedy weighted-utility ranking, sufficient when the candidate set is small or constraints decouple across configurations. The multi-model deployment study of Section~\ref{sec:eval-composability} demonstrates the ILP in action and contrasts it against an accuracy-greedy baseline that ignores deployment constraints.

%% file: sections/technical_impl.tex
\section{Technical Implementation}

HoliBench is implemented in approximately 15k lines of Python, structured around the four modular layers (Section~\ref{sec:architecture}). This section describes the implementation choices that realize that architecture.

\subsection{Configuration and Experiment Orchestration}
\label{subsec:config}

Experiments are configured using Hydra~\cite{yadan2019hydra}, a hierarchical configuration framework that composes YAML files at runtime. The configuration has three orthogonal axes which are the model (architecture, backend, quantization level), device (platform type, sampling intervals), and scenario (dataset, prompt template, sample count). These can easily be overridden from the CLI without modifying any files, allowing for modular and reproducible sweeps. Hydra also documents the configuration with every captured result.

\subsection{Hardware Profiling}
\label{subsec:hw_profiling}

Profilers are initialized by a base class, each implementing the three methods \texttt{start()}, \texttt{sample()}, and \texttt{stop()}. These profilers register themselves at import time with a name, priority, and availability predicate. At runtime, the synchronizer, after inspecting the host environment, instantiates only the profilers whose predicates are satisfied. This allows for the decoupling of platform support from the core framework. Hence, adding a new device to Holibench only requires implementing a single profiler class conforming to the interface contract without any modifications to the rest of the code.

\subsection{Model Loading}
\label{subsec:model_loading}

The model loader has two levels. The first level distinguishes the FM category (LLM, VLM, or TSFM) and then routes to the appropriate modality-specific driver. The second level identifies the inference engine (HuggingFace Transformers, vLLM, sglang, etc.) and then selects quantization and concurrency/batching configuration. All loaders have the same lifecycle with a \texttt{load}, \texttt{predict}, and then \texttt{unload} method. Backend-specific libraries are imported lazily only when the corresponding loader is instantiated to minimize startup overhead. Models are unloaded immediately after their scenario completes, freeing device memory for the next configuration. This release step is particularly important on memory-constrained devices such as the Raspberry Pi, where residual model weights from a prior run can prevent subsequent configurations from loading.

\subsection{Benchmark Suite}
\label{subsec:benchmark_suite}

Individual runs are orchestrated by a benchmark suite script that accepts a test matrix consisting of lists of models, scenarios, devices, and quantization levels, and executes every valid combination in sequence. Configurations that fail the pre-flight memory feasibility check are skipped automatically. After all runs complete, the suite aggregates the per-run \texttt{summary.json} files into a unified CSV, then generates comparative tables and figures for analysis.

\subsection{LLM-as-a-Judge}
\label{subsec:llm_judge}

Automatic string-matching metrics are insufficient for open-ended generation tasks such as summarization and visual question answering. HoliBench therefore includes an offline re-evaluation tool that replays model predictions through an LLM judge. The judge is prompted with a structured rubric covering exact match, semantic equivalence, hierarchical label matching, multiple-choice, and classification criteria. For the purposes of this work, we use Qwen2.5-32B-Instruct as the judge for our scenarios since the model has been found to be effective for such tasks \cite{sternlicht-etal-2025-debatable, yu-etal-2025-improve}.

\subsection{Configuration Selection Front-End}
\label{subsec:frontend}

The configuration-selection module described in Section~\ref{sec:optimizer} is exposed as a Flask web application providing a decision-support interface for users who need to deploy a model under resource constraints. Upon initiation, the application scans the results directory to build a merged table of latency, energy, and accuracy indexed by device, model, scenario, and quantization level. Users select a target device and scenario, adjust per-metric weights with sliders, and optionally impose hard constraints on latency, power, memory, or minimum accuracy. The solver runs on this filtered table and returns a ranked selection (model, batch size, inference engine, etc.) rendered alongside an interactive scatter plot of the feasible design space

%% file: sections/exp_setup.tex
\section{Experimental Setup}
\label{sec:exp}

\begin{table}[t]
\centering
\small
\setlength{\tabcolsep}{4pt}
\begin{tabular}{llll}
\hline
\textbf{Type} & \textbf{Task} & \textbf{Dataset} & \textbf{Metric} \\
\hline
\multirow{9}{*}{LLM}
 & NER                       & Wikiann~\cite{pan2017wikiann}           & F1       \\
 & Emotion classif.          & GoEmotions~\cite{demszky2020goemotions} & Acc.     \\
 & Sentiment classif.        & SST-2~\cite{socher2013sst}              & Acc.     \\
 &                           & IMDB~\cite{maas-etal-2011-learning}                &          \\
 & Topic classif.            & AG News~\cite{zhang2015character}          & Acc.     \\
 & Summarization             & CNN/DM~\cite{hermann2015cnndailymail}   & ROUGE-L  \\
 & Translation               & WMT16~\cite{bojar16}             & BLEU     \\
 & Code generation           & HumanEval~\cite{chen2021humaneval}      & pass@1   \\
 & Perplexity                & WikiText-2~\cite{merity2016pointer}    & PPL      \\
 &                           & C4~\cite{raffel2020exploring}                  &          \\
\hline
\multirow{9}{*}{VLM}
 & Object recognition        & CIFAR-100~\cite{krizhevsky2009cifar}    & Acc.     \\
 &                           & ImageNet~\cite{deng2009imagenet}        &          \\
 & Visual QA                 & VQAv2~\cite{goyal2017vqav2}             & Acc.     \\
 & Visual counting           & CountBenchQA~\cite{beyer2024paligemma},  & Acc.   \\
 &                           & \cite{paiss2023countclip}       &        \\
 & Document QA               & DocVQA~\cite{mathew2021docvqa}          & Acc.     \\
 & Traffic sign classif.     & GTSRB~\cite{stallkamp2012gtsrb}         & Acc.     \\
 & Gesture recognition       & HaGRID~\cite{Kapitanov_2024_WACV}       & Acc.     \\
 & Gender recognition        & UTKFace~\cite{zhifei2017cvpr}        & Acc.     \\
 & Location recognition      & Country211~\cite{radford2021learning}       & Acc.     \\
\hline
\multirow{3}{*}{TSFM}
 & Outlier detection         & NAB~\cite{lavin2015nab}                 & F1       \\
 & Changepoint detection     & GIFT-EVAL~\cite{aksu2024gifteval}       & F1       \\
 & Forecasting               & M3 Monthly~\cite{makridakis2000m3}      & SMAPE    \\
\hline
\end{tabular}
\caption{Evaluation tasks. Each task is encapsulated as a modality-specific scenario with its own dataset, input formatting, and scoring metric.}
\vspace{-1em}
\label{tab:tasks}
\end{table}

The evaluation of HoliBench is designed to exercise the PAL (Section~\ref{sec:pal}) and the modality-aware drivers across representative device, edge, and cloud platforms. We construct a heterogeneous testbed that mirrors the structural challenges identified in Section~\ref{sec:intro}, i.e.,  disparate telemetry interfaces, varying compute architectures, and diverse FM workloads.

\subsection{Hardware Testbed}
\label{sec:exp:hw}

Table~\ref{tab:platforms} summarizes the seven platforms spanning four deployment tiers, each exercising a distinct profiler backend in the platform abstraction layer.
In the device tier, a Raspberry Pi~5 represents power-constrained single-board computers common in IoT sensing and low-power monitoring. At the edge tier, the NVIDIA Jetson Orin AGX provides GPU-accelerated inference within the power envelope typical of robotics and autonomous systems, while the RTX~3070 offers a consumer-grade Ampere GPU with Tensor Core support for low-precision arithmetic. At the cloud tier, the RTX~5070 (Blackwell) and RTX~A5000 (Ampere) represent professional and next-generation discrete GPUs with larger VRAM budgets. The Mac Mini M2 and Mac Mini M4 constitute the workstation tier and introduce Apple Silicon's unified memory architecture, against which discrete-GPU and ARM SBC platforms can be compared.

Each platform exposes a fundamentally different telemetry interface for power measurement. The PAL normalizes these into consistent energy and utilization metrics, and verifying that this normalization produces commensurable results across backends is itself a goal of our evaluation.

\subsection{Model and Task Selection}
\label{sec:exp:models}

Following the modality-aware architecture (Section~\ref{sec:architecture}), we select models representing the three workload classes HoliBench supports. While the current evaluation covers LLMs, VLMs, and TSFMs, the framework's modality-driver architecture extends to additional modalities such as audio through the same interface contracts.

For LLMs, we evaluate models from the SmolLM~\cite{allal2025smollm2} (135M to 3B), Llama-3.2~\cite{llama3} (1B to 3B), Qwen2.5~\cite{qwen25} and Qwen3~\cite{qwen3} (0.6B to 8B), Gemma~\cite{gemma3team2025} (1B to 4B), and DeepSeek~\cite{deepseek2025r1} (7B to 8B) families, spanning the full range of parameter counts deployable on the hardware testbed. Each model is evaluated at native precision and, where the backend supports it, under INT8 and INT4 quantization. For VLMs, we evaluate SmolVLM~\cite{marafioti2025smolvlm} (256M to 1.7B), Qwen2.5-VL~\cite{qwen25vl2024} (3B), PaLIGemma~\cite{beyer2024paligemma} (3B), and Moondream~\cite{moondream} (1.6B), which exercise the phase-level decomposition between vision encoding and language generation described in Section~\ref{sec:timing-model}. For TSFMs, we evaluate Chronos~\cite{ansari2024chronos} (8M to 46M), MOMENT~\cite{goswami2024moment} (385M), and Granite-TS~\cite{granitets2024} (2M), whose single-pass windowed inference path produces hardware utilization patterns distinct from autoregressive models.

Table~\ref{tab:tasks} lists the evaluation scenarios with their associated datasets and performance metrics, spanning classification, generation, and regression tasks across all three modalities. Each scenario executes 100 evaluation samples per model-device-quantization setting.

\subsection{Measurement Protocol}
\label{sec:exp:protocol}

Beyond the inference timing strategies defined in Section~\ref{sec:timing-model}, several experimental parameters govern data collection. The power sampling interval is configured per modality: 100\,ms for LLM and VLM workloads where inference typically spans multiple seconds, and 10\,ms for TSFM workloads where single-pass inference completes in tens of milliseconds. Each scenario executes 100 samples per model-device-quantization configuration. All profiling uses single-query sequential inference, reflecting the request-at-a-time pattern characteristic of on-device CPS deployment. The full experimentation sweep crosses the model set, up to three quantization levels (FP16, INT8, INT4), and all seven platforms, with the pre-flight feasibility filter reducing the configuration count on memory-constrained devices.

To confirm that HoliBench's accuracy pipeline introduces no measurement artifacts, we validate against EleutherAI's Model Evaluation Harness on overlapping models and tasks.

%% file: sections/evaluation.tex
\section{Evaluation}

\begin{figure}[t!]
    \centering
    \includegraphics[width=0.90\columnwidth]{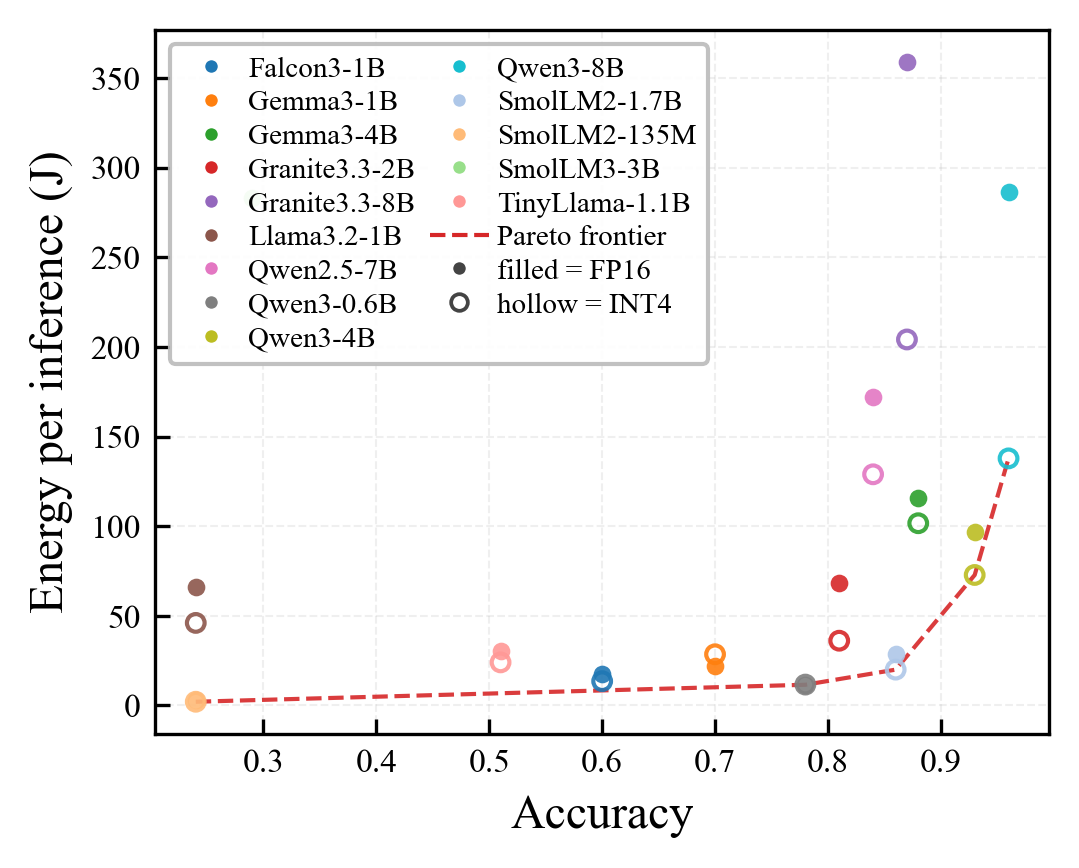}
    \caption{Normalized accuracy vs.\ net inference energy for the summarization
    task on the A5000. The Pareto frontier (dashed)
    traces the minimum energy required at each accuracy level.}
    \vspace{-1em}
    \label{fig:acc}
\end{figure}

\begin{figure*}[t]
    \centering
    \begin{subfigure}[t]{0.40\textwidth}
        \centering
        \includegraphics[width=\textwidth]{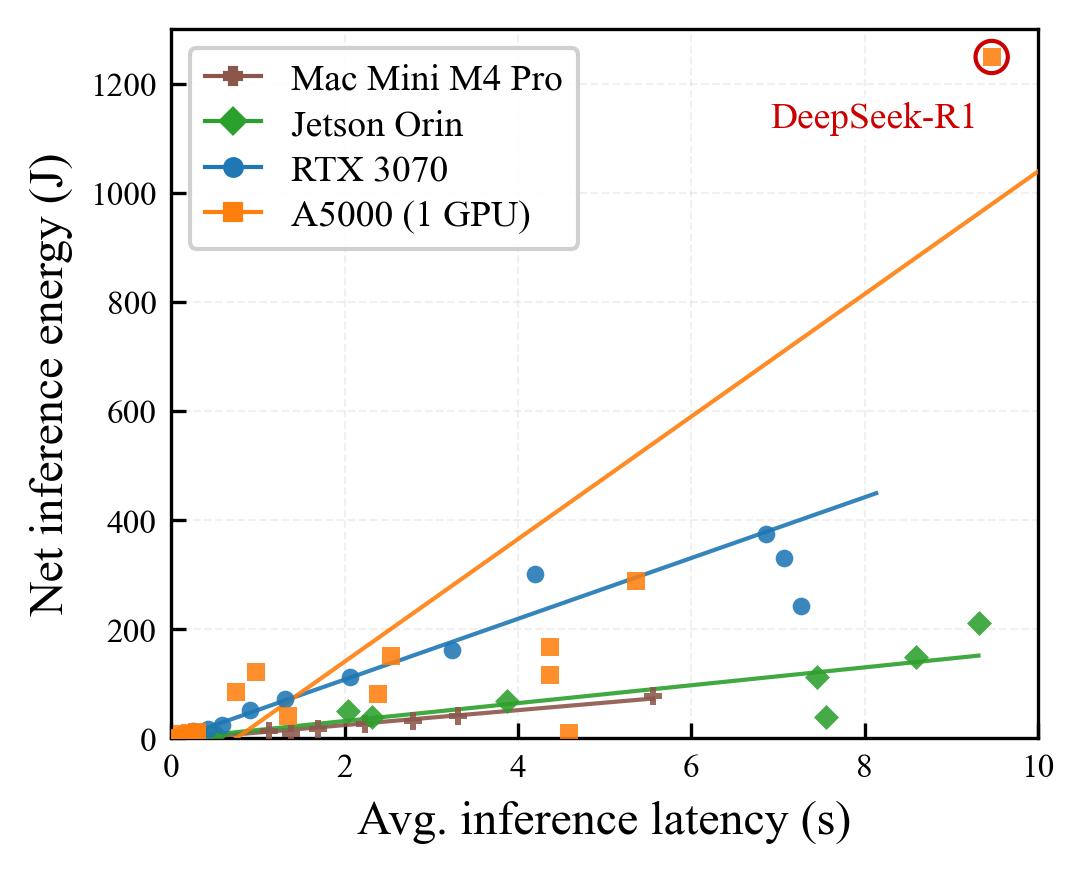}
        \caption{}
        \label{fig:char-energy}
    \end{subfigure}
    \hspace{3em}%
    \begin{subfigure}[t]{0.45\textwidth}
        \centering
        \includegraphics[width=\textwidth]{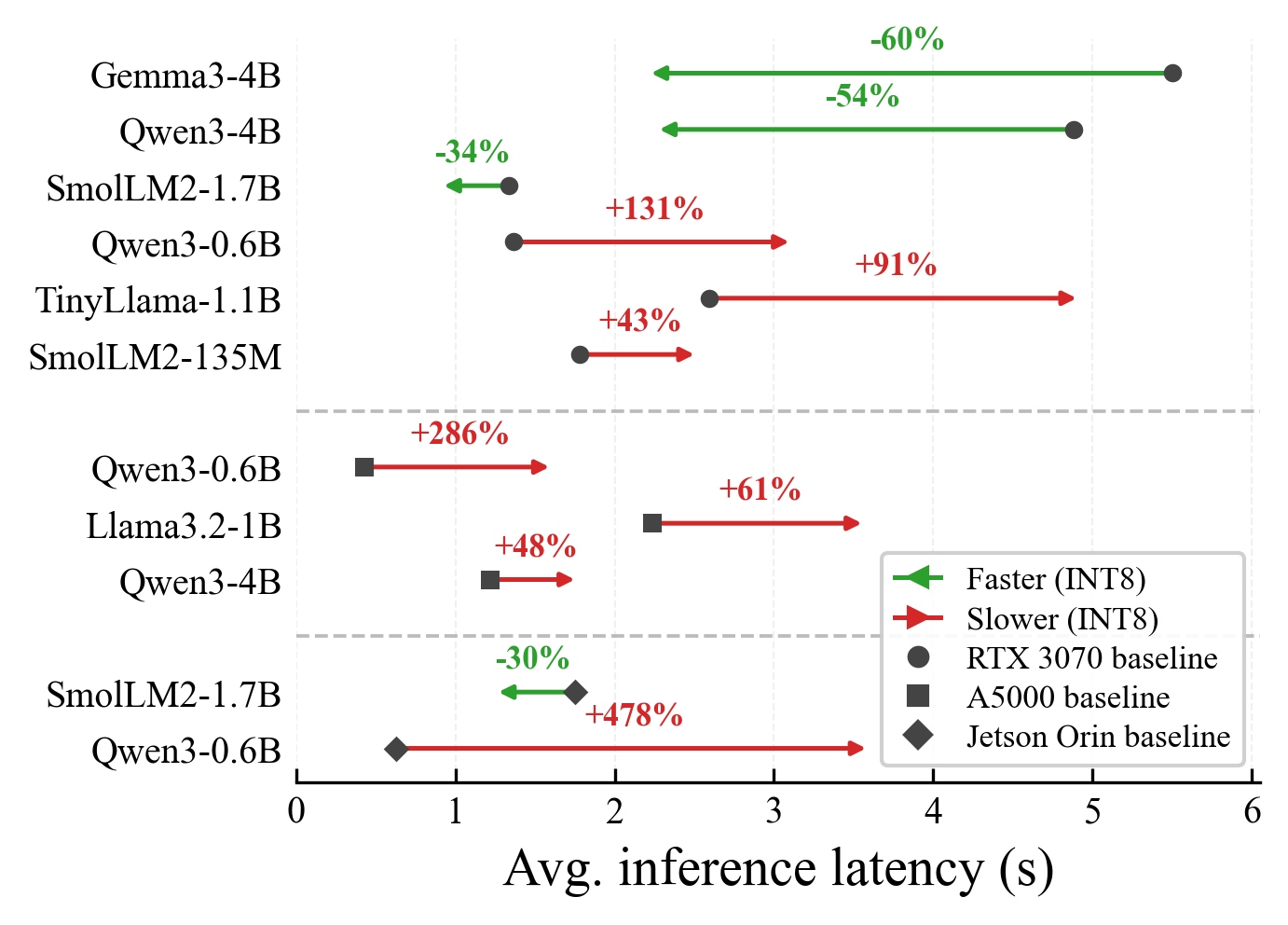}
        \caption{}
        \label{fig:char-quantization}
    \end{subfigure}
    \caption{\textbf{(a)}~Net inference energy vs.\ latency. Per-device regression
    slopes correspond to average inference power,
    confirming that power draw is approximately device-determined.
    \textbf{(b)}~Impact of INT4 weight-only quantization on inference latency across three hardware platforms. Each arrow connects a model's FP16 baseline (circle) to its INT4 latency; green arrows pointing left indicate speedup, red arrows pointing right indicate a dequantization penalty.}
    \label{fig:characterization}
\end{figure*}

We use HoliBench to surface deployment tradeoffs across the profiled design space that are invisible to accuracy-only evaluation. The results are organized around three characterization findings, each revealing a dimension of the deployment decision that existing benchmarking frameworks do not capture, followed by a cross-platform validation of the PAL.

\subsection{Diminishing Accuracy Returns Under Energy Constraints}
\label{subsec:fig2}
Figure~\ref{fig:acc} plots accuracy against the net inference energy for 15 LLM configurations on the summarization task, evaluated on the A5000. The Pareto frontier traces the minimum energy required to achieve each accuracy level and reveals a characteristic convex shape. In the low-to-moderate accuracy range, efficient configurations exist at negligible energy cost, but beyond approximately 80\,\% accuracy the frontier inflects sharply, with each incremental accuracy gain demanding a disproportionately larger energy budget. The ratio of energy cost per accuracy point in the upper range exceeds that in the lower range by over an order of magnitude.

Equally significant is the vertical spread above the frontier. At comparable accuracy levels, configurations differ by as much as 12-fold in energy consumption. These are functionally equivalent models with vastly different deployment costs. A user selecting by accuracy alone would have no basis for distinguishing between them, as the energy dimension is invisible without hardware-in-the-loop profiling. HoliBench surfaces this dimension, and the selection interface exposes these Pareto-efficient configurations under user-specified constraints, as demonstrated in the multi-model deployment study of Section~\ref{sec:eval-composability}.

\subsection{Latency-Energy Linearity and Predictive Power Estimation}
\label{subsec:fig3}
Figure~\ref{fig:char-energy} plots net inference energy against average inference latency for a classification task across four devices spanning the edge-to-server continuum. On every device tested, energy scales linearly with latency. The per-device regression slopes recover the average inference power during autoregressive decode, with the A5000 drawing approximately twice the power of the RTX 3070 and the Jetson AGX Orin and Mac Mini M4 Pro each operating at roughly an order of magnitude below either discrete GPU. This linearity follows from the memory-bandwidth-bound nature of autoregressive decoding, where each decode step is dominated by weight transfers from device memory and the arithmetic intensity is too low to saturate the GPU's compute units. Power draw is therefore set primarily by the memory subsystem's operating envelope rather than by model size. The Mac Mini M4 Pro achieves power draw comparable to the Jetson despite delivering higher decode throughput because its unified memory architecture eliminates the discrete host-to-device transfers that CUDA platforms require for every weight fetch.

The fit holds from sub-second inference on 135M-parameter models through multi-second runs on 8B-parameter models. One configuration deviates visibly: DeepSeek-R1-Distill-Llama-8B (circled), a chain-of-thought reasoning model that emits several hundred intermediate reasoning tokens for a task requiring a single classification label. The resulting output inflation places this configuration well above the regression line. This is a deployment pathology, not a model quality issue, as the model's classification accuracy may remain competitive while its per-inference energy cost exceeds that of comparably-sized models by an order of magnitude.

The key regularity is that average inference power is approximately stable for a given device regardless of model or workload, enabling predictive energy estimation from latency measurements alone.

\begin{figure}[t]
    \centering
    \includegraphics[width=0.95\columnwidth]{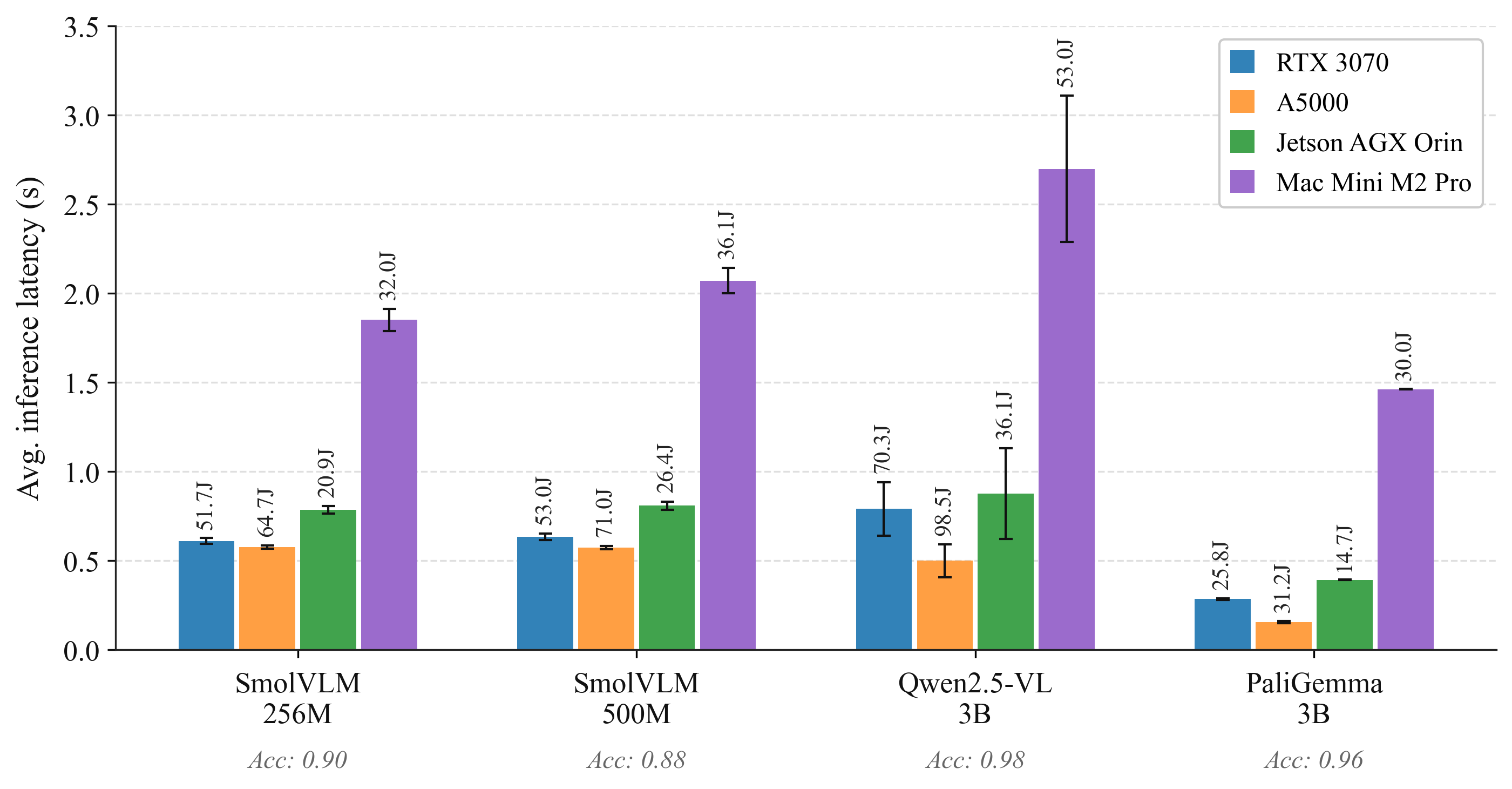}
    \caption{Cross-platform comparison of four VLMs on a gender recognition task, showing inference latency and net energy per inference across four hardware platforms. Error bars represent the standard error of the mean latency.}
    \vspace{-1em}
    \label{fig:char-pal}
\end{figure}

\subsection{Hardware-Dependent Quantization Effects}
\label{subsec:quant}
Figure~\ref{fig:char-quantization} evaluates INT4 weight-only quantization (bitsandbytes) on the HuggingFace inference backend across three hardware platforms. The common expectation that quantization uniformly reduces inference latency does not hold under these conditions. On the RTX 3070, quantization benefits are strongly model-size-dependent. The 4B-parameter models achieve 54--60\,\% latency reductions, consistent with larger weight matrices shifting the bottleneck from compute to memory transfer, where INT4 compression halves the transfer volume. For sub-1B models, however, quantization increases latency by 43--131\,\%, because the weight matrices are small enough that memory bandwidth is not saturated at FP16 and the dequantization overhead of converting INT4 weights back to FP16 on every forward pass exceeds the transfer savings.
On the A5000, quantization increases latency across all tested models, including Qwen3-4B, which achieved a 54\,\% speedup on the RTX 3070 but slows down by 48\,\% on the A5000. Both GPUs are Ampere-generation with Tensor Core support. The divergence stems from the bitsandbytes INT4 kernel implementation, which does not use the same optimized code paths on professional-class (A-series) GPUs as on consumer-class (GeForce) hardware.

On the Jetson AGX Orin, the pattern is similar. Qwen3-0.6B incurs a 478\,\% latency penalty under INT4, while SmolLM2-1.7B is one of the few configurations that benefits. These results illustrate a three-way interaction between model size, quantization runtime, and hardware microarchitecture that cannot be predicted from accuracy benchmarks or hardware specifications alone. A user following the common heuristic that ``quantization is always faster'' would degrade deployment performance on the majority of configurations tested here. HoliBench exposes this interaction by profiling the actual quantized execution path on the target device.

\begin{figure*}[t]
    \centering
    \begin{subfigure}[t]{0.40\textwidth}
        \centering
        \includegraphics[width=\textwidth]{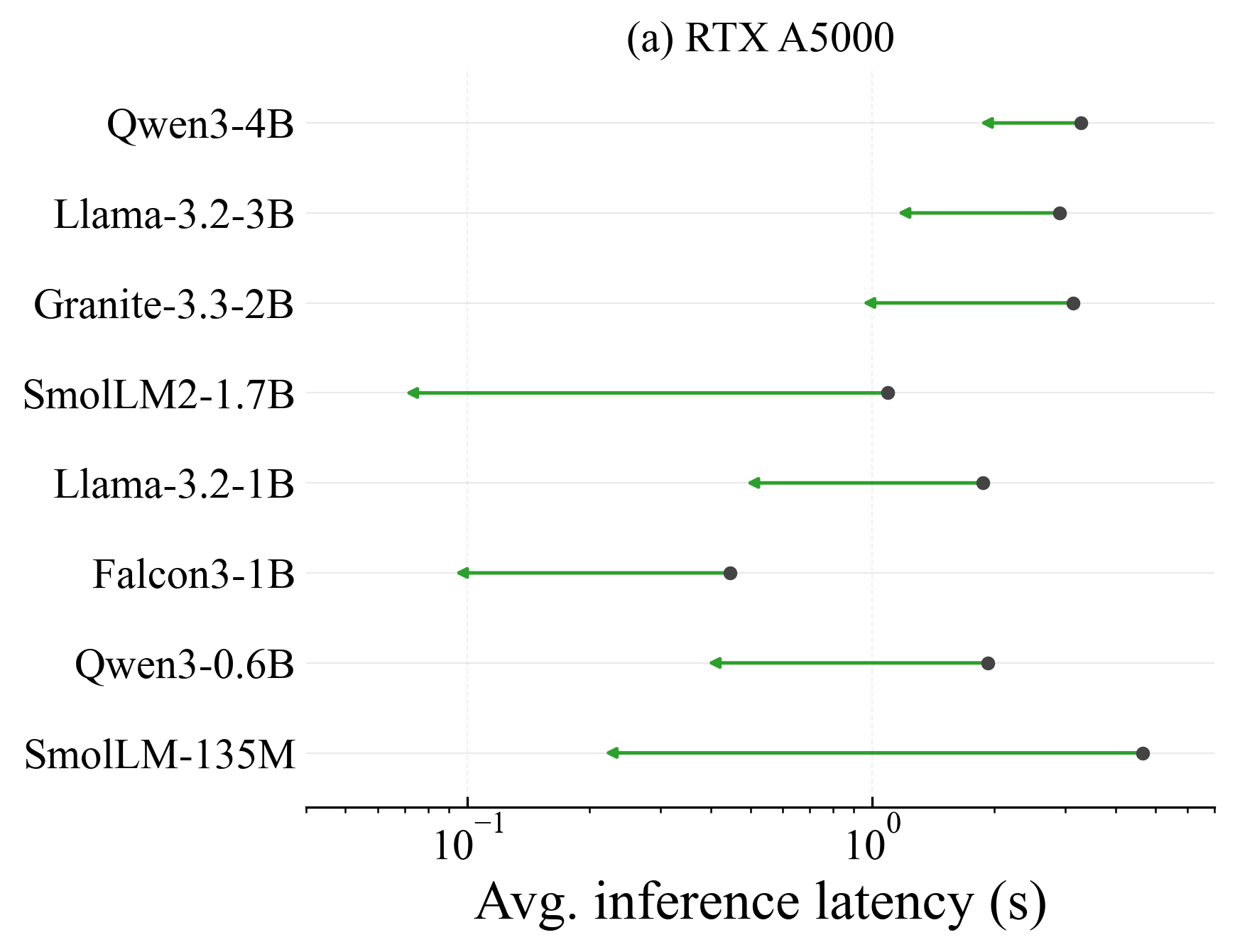}
        \caption{}
        \label{fig:a5000}
    \end{subfigure}
    \hspace{3em}%
    \begin{subfigure}[t]{0.40\textwidth}
        \centering
        \includegraphics[width=\textwidth]{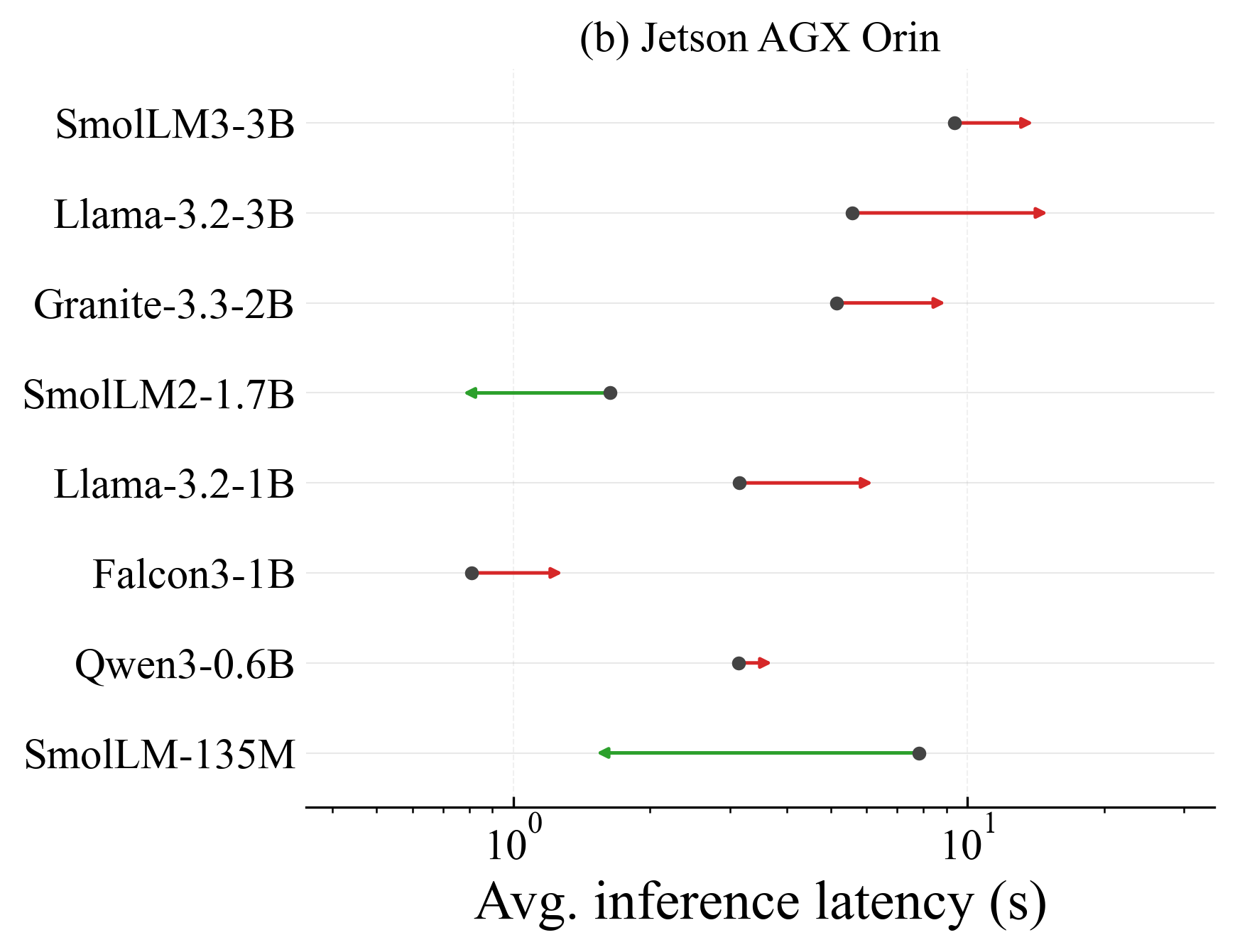}
        \caption{}
        \label{fig:jetson}
    \end{subfigure}
    \caption{Latency shift under vLLM relative to HuggingFace Transformers across a representative LLM set on the \textbf{(a)}~RTX~A5000 and \textbf{(b)}~Jetson~AGX~Orin. Each arrow connects a model's HuggingFace baseline (circle) to its vLLM latency. \textcolor{green}{green} leftward arrows indicate vLLM is faster, \textcolor{red}{red} rightward arrows indicate vLLM is slower. Models are sorted by parameter count.}
    \label{fig:backend}
\end{figure*}

\subsection{Cross-Platform Measurement Validation}
\label{subsec:fig5}
Figure~\ref{fig:char-pal} compares four vision-language models on a gender recognition task across four hardware platforms, validating that HoliBench's PAL produces commensurable measurements from heterogeneous telemetry sources (NVML on discrete CUDA GPUs, jtop on the Jetson AGX Orin, and powermetrics on Apple Silicon). Task accuracy is stable for each model across platforms, confirming that the profiling infrastructure introduces no computational artifacts.

Deployment cost, by contrast, varies substantially even for the same model. PaliGemma-3B completes inference in 0.15\,s on the A5000 but consumes 31.2\,J, while the Jetson takes 2.6 times longer at 0.39\,s but uses less than half the energy at 14.7\,J. This inversion between latency ranking and energy ranking is consistent across all four models and follows directly from the power characteristics of each platform. The A5000's high memory bandwidth minimizes inference time but its active power draw exceeds that of the Jetson by roughly an order of magnitude, so the energy product is higher despite shorter execution. The Jetson, operating within a constrained power envelope designed for edge deployment, trades latency for efficiency. The Mac Mini M2 Pro incurs higher latency during inference due to the overhead of standard HuggingFace implementations lacking native MLX acceleration, but its power efficiency keeps its cumulative energy cost largely comparable to, and often below, that of the other platforms. These measurements originate from four distinct telemetry backends, each with different sampling rates and access interfaces. A quantitative comparison is possible only because the PAL normalizes them into a common metric.

A user optimizing for battery life on a mobile platform would select the Jetson, while one optimizing for throughput on a server would select the A5000. These are opposing deployment decisions for the same model, and neither conclusion is reachable without cross-platform energy profiling that HoliBench provides.

\subsection{Cross-Backend Generalization}
\label{sec:eval:backend}

\begin{figure*}[t]
    \centering
    \begin{subfigure}[t]{0.48\textwidth}
        \centering
        \includegraphics[width=\textwidth]{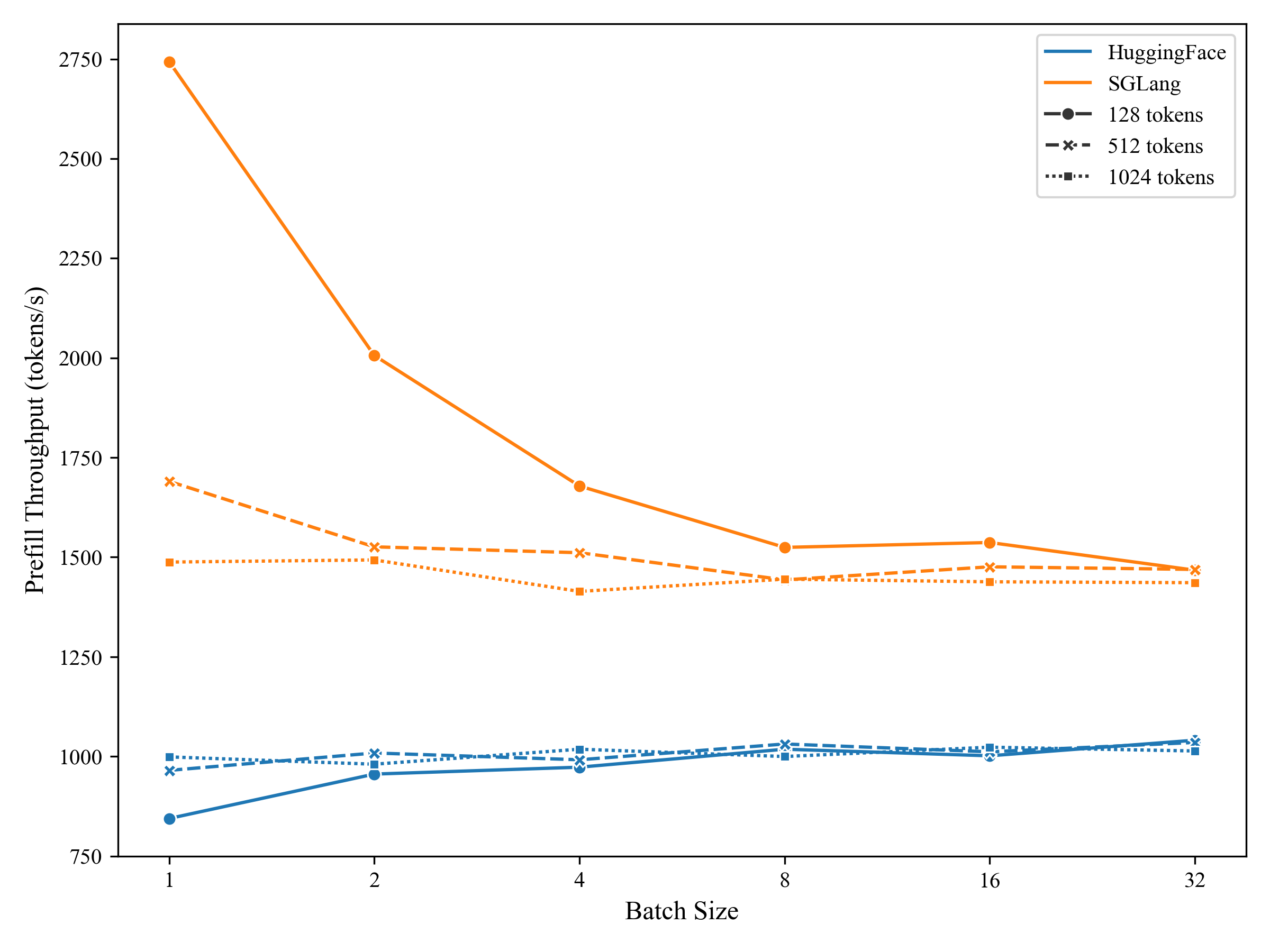}
        \caption{Prefill Throughput vs Batch Size}
        \label{fig:prefill}
    \end{subfigure}
    \hspace{1em}%
    \begin{subfigure}[t]{0.48\textwidth}
        \centering
        \includegraphics[width=\textwidth]{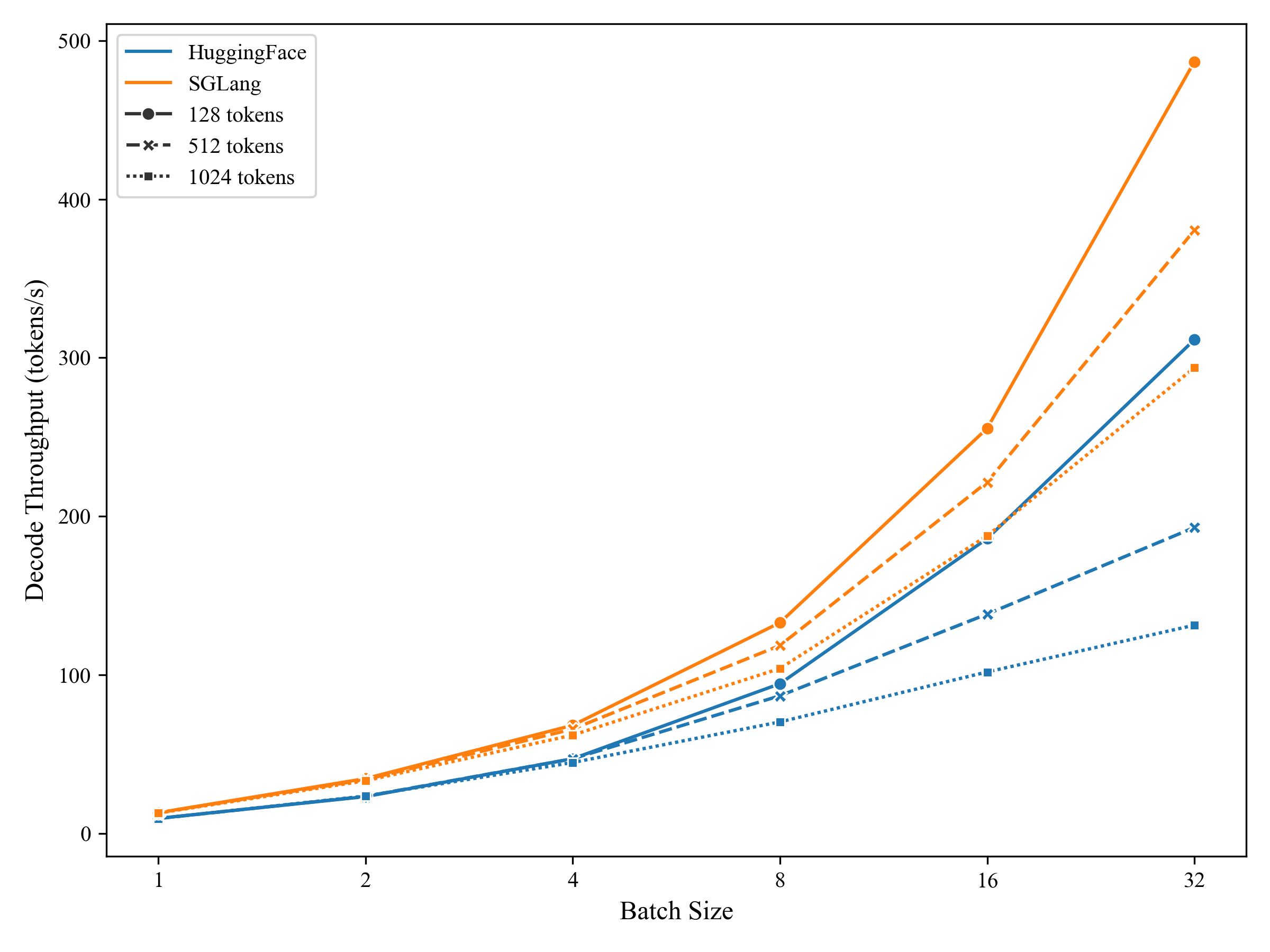}
        \caption{Decode Throughput vs Batch Size}
        \label{fig:decode}
    \end{subfigure}
    \caption{Comparison of SGLang vs HuggingFace across differing batch sizes and sequence lengths to demonstrate advantages of different inference engines across different configurations}
    \label{fig:batch-fig}
\end{figure*}

All preceding characterization uses HuggingFace Transformers, which is widely adopted for prototyping and initial edge deployment. This subsection addresses two questions: (1) do the characterization findings of Sections~\ref{subsec:fig2} through~\ref{subsec:fig5} persist across inference backends, and (2) can new backends be integrated through HoliBench's existing interface contracts? We re-profile a representative subset of LLMs spanning 135M to 4B parameters on the RTX~A5000 and Jetson~AGX~Orin under vLLM, TensorRT-LLM, SGLang, and HuggingFace. No modifications to the PAL, modality-aware driver, or orchestration logic were required to add either backend.

The energy-latency linearity identified in Section~\ref{subsec:fig3} persists across all evaluated backends (R$^2 > 0.98$ under vLLM, TensorRT-LLM, and SGLang compared to $0.82$--$0.86$ under HuggingFace), indicating that the underlying model-hardware relationship is not specific to HuggingFace execution. Figure~\ref{fig:backend} compares latency under HuggingFace and vLLM. On the RTX~A5000, vLLM reduces latency for every model evaluated, with speedups ranging from 1.8$\times$ to 16$\times$. On the Jetson~AGX~Orin, however, the advantage does not hold uniformly. All models at 2B parameters and above run slower under vLLM, with Llama-3.2-3B exhibiting the largest slowdown at 2.8$\times$. Smaller models show mixed behavior, with some benefiting from vLLM and others slowing down. One plausible explanation is that vLLM's serving-oriented machinery,
including paged KV-cache management, request scheduling, and
continuous-batching infrastructure, introduces overheads that
amortize effectively in high-throughput serving environments but
not under the single-query inference workloads evaluated on the
Jetson. The fact that the slowdown appears primarily for larger
models suggests that these overheads interact with model scale,
though isolating the precise mechanism is beyond the scope of this
study.

The same runtime therefore produces opposite deployment outcomes across hardware platforms. A user selecting a runtime from documentation alone would not predict this inversion; hardware-in-the-loop profiling is necessary to surface it. TensorRT-LLM and SGLang on the A5000 exhibit behavior similar to vLLM and preserves the energy-latency trends observed throughout the paper. Overall, the key findings of Sections~\ref{subsec:fig2} through~\ref{subsec:fig5} persist across inference backends, while absolute performance remains strongly dependent on runtime choice. HoliBench's profile contract accommodates these backends through the existing driver interface, enabling deployment decisions to be made empirically rather than from runtime-level documentation alone.

We present a comparative empirical evaluation in Figure 8, benchmarking SGLang against a baseline HuggingFace deployment on Falcon3-1b across a sweep of sequence lengths ($L \in \{128, 512, 1024\}$) and batch sizes ($B \in \{1, 2, 4, 8, 16, 32\}$). Our contributions derived from this case study are twofold. First, we quantitatively demonstrates the performance imperative of accommodating specialized inference backends. On Falon3-1B, SGLang exhibits strictly superior prefill and decode throughputs, an advantage fundamentally attributable to its integration of hardware-optimized custom execution kernels and radix attention mechanisms for efficient KV-cache state sharing. Second, leveraging HoliBench’s profiling telemetry, we expose the underlying scaling dynamics of various batching configurations. At lower degrees of concurrency and shorter context windows, SGLang does better. However, as concurrency scales, the relative prefill throughput advantage of SGLang asymptotically converges to a constant factor of approximately 1.5$\times$. Furthermore, our profiling highlights divergent scaling behaviors between operational phases. Because the prefill phase is inherently compute-bound and already highly parallelized, increasing the batch size yields diminishing returns. Conversely, batching provides substantial throughput scaling during the auto-regressive decode phase by amortizing memory bandwidth costs. Ultimately, the framework's optimizer uses such collected data along with the aforementioned large-model degradation analysis to recommend optimal deployment configurations, encompassing both inference engine selection and batching strategy.

%% file: sections/case_study.tex
\section{Case Study: Profile Composability for Multi-Model CPS Deployment}
\label{sec:eval-composability}

\begin{figure*}[t]
    \centering
    \begin{subfigure}[t]{0.50\textwidth}
        \centering
        \includegraphics[width=\textwidth]{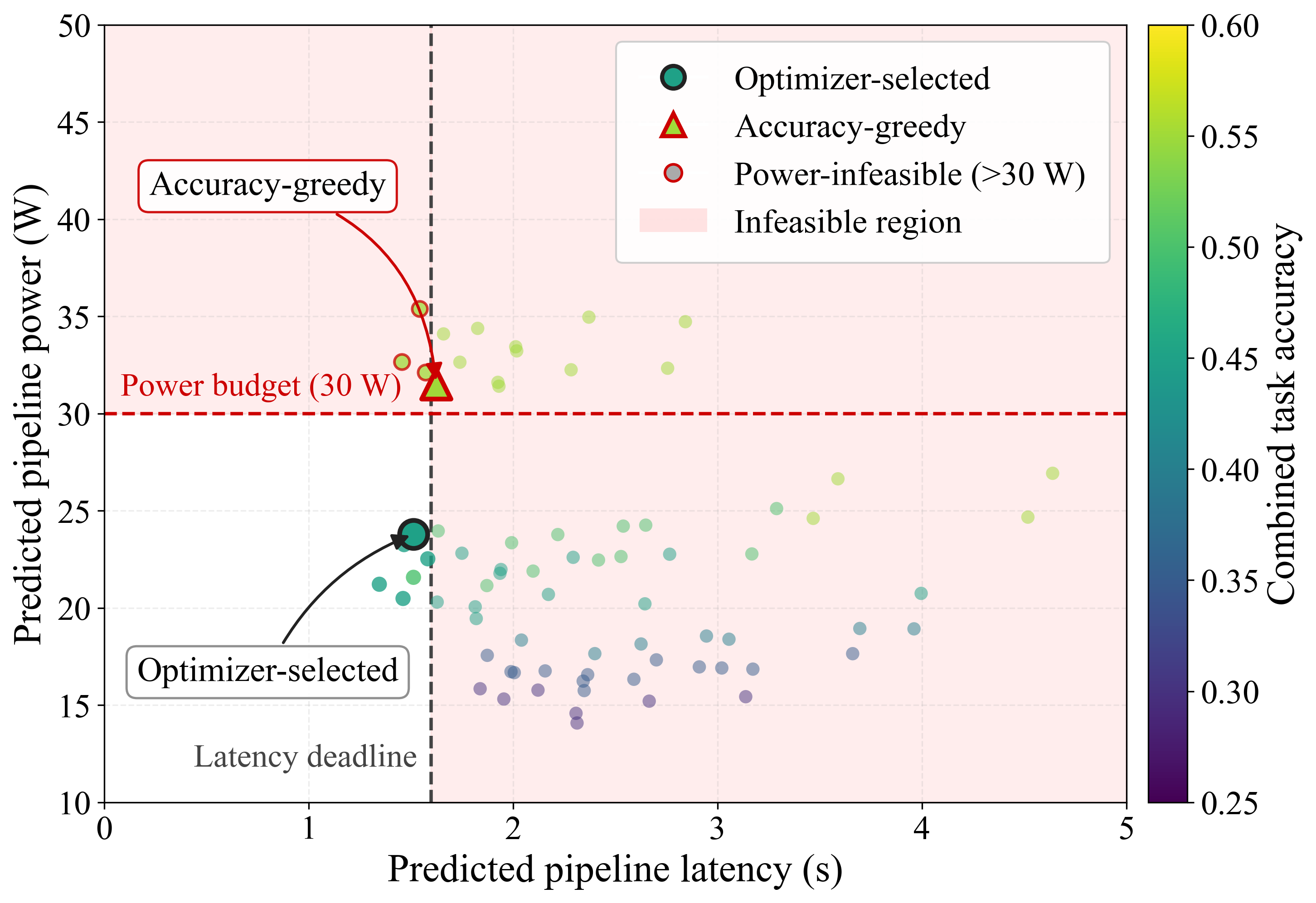}
        \caption{}
        \label{fig:composability-design}
    \end{subfigure}
    \hspace{1em}
    \begin{subfigure}[t]{0.45\textwidth}
        \centering
        \includegraphics[width=\textwidth]{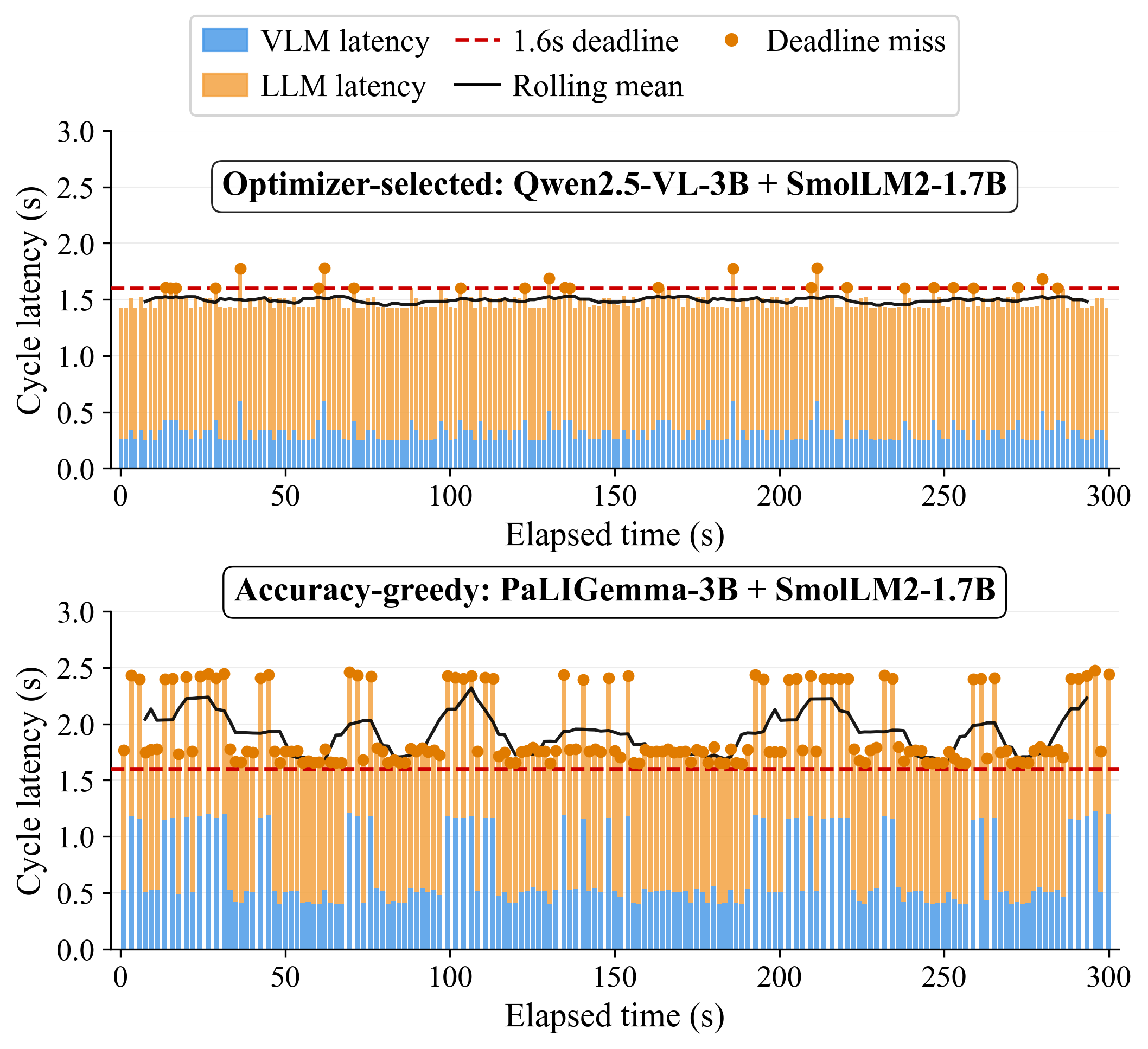}
        \caption{}
        \label{fig:composability-validation}
    \end{subfigure}
    \caption{Profile composability for multi-model deployment on the Jetson AGX Orin 64~GB.
    \textbf{(a)}~Design space of (VLM, LLM) pairings by predicted pipeline latency vs.\ predicted pipeline power, with color encoding combined task accuracy. The feasible region is shaded in red from where the solver selects the highest-accuracy feasible pairing (circled). The accuracy-greedy selection (triangle) violates both constraints.
    \textbf{(b)}~Per-cycle latency over a 10-minute co-resident CARLA run, decomposed into VLM (blue) and LLM (orange) contributions. The solver-selected configuration holds below the deadline on 88.8\,\% of cycles while the accuracy-greedy configuration exceeds the deadline on every cycle.}
    \vspace{-0.5em}
    \label{fig:composability}
\end{figure*}

The preceding experiments evaluate individual model-device-quant. configurations in isolation.
In practice, CPS applications frequently require multiple foundation models to share a single device,
with each model contributing to a different stage of the processing pipeline.
A vision-language model processing camera frames for scene understanding may feed its output
to a language model responsible for high-level planning, with both models resident in device memory
and executing sequentially within each control cycle.
Whether HoliBench's single-model profiling data, collected when each model has exclusive access
to the device, remains predictive when models share hardware resources is not guaranteed.
Co-resident model weights increase baseline memory pressure,
and sequential execution of heterogeneous workloads (compute-bound vision encoding followed
by bandwidth-bound autoregressive decode) may interact with memory bandwidth
in ways that isolated profiling does not capture.

We evaluate this with a dual-FM perception-planning pipeline deployed in the CARLA urban driving
simulator~\cite{dosovitskiy2017carla}, with FM inference executing on the NVIDIA Jetson AGX Orin
64\,GB, the same class of edge compute module deployed in production autonomous systems including the Serve Robotics sidewalk delivery fleet~\cite{serverobotics2024gen3} and Toyota's DRIVE AGX platform~\cite{nvidia_drive_agx}.
An autonomous service robot navigates CARLA Town~03 at approximately 3\,mph, a representative
operating speed for sidewalk delivery platforms, requiring each perception-planning cycle to complete within 1.6\,s.
Sustained inference power is budgeted at 30\,W to preserve battery life.
All candidate VLM + LLM pairings fit within the Orin's 64\,GB unified memory, which may lead
a user to conclude that any configuration is deployable.
The binding constraints are latency and sustained power draw under continuous operation,
neither of which is captured by accuracy-only evaluation tools.

\subsection{Estimating deployment cost without task-specific profiling}
The VLM performs image captioning on each camera frame, for which standard benchmark datasets
exist and HoliBench has direct profiling data.
The LLM produces a short structured navigation command of approximately 25 tokens, a task
with no corresponding benchmark dataset.
Typically, evaluating such a task would require constructing a dedicated evaluation set and
profiling the pipeline from scratch.
We show that HoliBench's existing profiling data and the insights derived from the preceding
characterization are sufficient to estimate deployment cost without task-specific re-profiling.

For latency, we exploit the linear relationship between output token count and autoregressive
decode time, a well-established property of transformer
inference~\cite{pope2023efficiently}.
HoliBench's modality-aware drivers decompose each LLM inference into prefill and decode phases,
exposing the per-token decode cost directly.
We extract this cost from profiled classification and summarization endpoints and interpolate
to the target 25-token output length.

Latency interpolation alone does not resolve the power dimension.
We lack power measurements for the specific 25-token generation task, and power draw need not
be constant across output lengths since prefill is compute-bound and may draw higher wattage than
the memory-bandwidth-bound decode phase.
However, the characterization results in Section~\ref{subsec:fig3} establish that
for a given device, average inference power is approximately stable
across output lengths.
We verify this for the candidate LLMs and find that across classification and summarization
tasks, average power deviates by less than 16\,\% for 9 of 10 configurations.
Given this near-constant power property, the power draw measured at any profiled output length transfers to the interpolated task.
Pipeline power is then computed as a time-weighted average of each model's standalone power,
weighted by the fraction of cycle time each model occupies. VLM latency and power are taken directly from profiled image captioning data, since the vision encoder's prefill cost dominates VLM inference and is independent of output length.

\subsection{Configuration selection}
Figure~\ref{fig:composability}(a) shows the candidate design space, with each point representing
a VLM + LLM pairing plotted by predicted combined latency against combined task accuracy,
color-coded by predicted pipeline power.
Since all models fit in memory, a user relying on accuracy benchmarks alone would select
PaLIGemma-3b paired with SmolLM2-1.7B, the highest-accuracy pairing.
This configuration exceeds the 30~W power budget and would reduce battery-powered mission
duration by approximately 24\,\% relative to a power-feasible alternative.

HoliBench's solver jointly constrains accuracy, latency, and power.
It selects Qwen2.5-VL-3B paired with SmolLM2-1.7B, satisfying both the latency deadline and
the power budget while accepting lower VLM accuracy.
This tradeoff is invisible to any evaluation framework that does not jointly profile latency,
power, and accuracy across configurations.

\subsection{Runtime validation}
Both configurations are loaded co-resident in GPU memory and executed sequentially over
10-minute CARLA runs.
Figure~\ref{fig:composability}(b) plots per-cycle latency over time.

The solver-selected configuration tracks its predicted latency within 1.2\,\% and its
predicted power within 2.5\,\%.
The LLM contributes 79\,\% of cycle time with 5.3~ms standard deviation, consistent with the
deterministic cost of fixed-length autoregressive decode.
Despite being the smaller model, the LLM dominates cycle time because 25 sequential decode
steps are inherently more expensive than the VLM's single-pass vision encoding.
This split is counterintuitive from a parameter count perspective and is only visible through
HoliBench's phase-level decomposition.
The VLM contributes the remaining 21\,\% with higher variance (std 74.5~ms), driven by
input-dependent vision encoder workload.
At the 1.6\,s deadline, 88.8\,\% of cycles comply, and all cycles complete within 1.8\,s.
The tail violations are attributable entirely to VLM input variance, with latency remaining
stable to within 10\,ms over the full 10-minute run. No co-residency power overhead is detectable. The idle model's VRAM footprint adds no measurable power draw beyond the system baseline already captured in standalone profiling.

The accuracy-greedy configuration exceeds the 1.6\,s deadline on every cycle
(mean 1896\,ms, worst case 2849\,ms) and draws 32.0\,W sustained, violating the power budget.
Its latency prediction error is $+$16.8\,\%, driven by PaLIGemma's inconsistent output behavior.
On 24\,\% of input frames, the model produces unexpectedly verbose outputs, inflating VLM latency
well beyond the profiled mean.
Standard benchmarks with curated inputs do not expose this behavior,
but it appears immediately under the sustained, diverse-input conditions typical of CPS deployment.

Sequential execution, static GPU memory allocation, and temporal separation of compute phases
underpin the composability observed here.
These conditions may not hold under concurrent execution or dynamic batching, where models
contend for shared compute and memory bandwidth.

%% file: sections/conclusion.tex
\section{Discussion and Conclusion}

\textit{The Dequantization Gap.}
Our quantization characterization (Section~\ref{subsec:quant}) exposes a gap between compression research and deployment reality. INT4 and INT8 quantization are widely treated as straightforward latency and memory optimizations, but our cross-platform measurements show that the benefit is contingent on hardware support and kernel implementation. On platforms lacking dedicated low-precision arithmetic units, or where the quantization runtime does not use optimized code paths for the target GPU class, dequantization overhead during inference offsets or exceeds the memory bandwidth savings that reduced bit-width provides. The divergence between the RTX 3070 and A5000, both Ampere-generation GPUs with Tensor Core support, illustrates that hardware generation alone does not predict quantization behavior. Quantization decisions require hardware-in-the-loop profiling because the interaction between model size, quantization runtime, and device microarchitecture cannot be predicted from accuracy-loss tables or hardware specifications alone.

\textit{Profile Composability and CPS Deployment.}
The case study (Section~\ref{sec:eval-composability}) assesses whether single-model profiles collected when each model has exclusive access to the device remain predictive when models share hardware resources. Under sequential co-resident execution on the embedded Jetson device, the solver-selected configuration tracks its predicted latency within 1.2\% and its predicted power within 2.5\%, with no detectable co-residency overhead. The time-weighted power model reduces pipeline energy estimation to a weighted sum of standalone measurements, and latency interpolation from phase-level decomposition extends the profiled design space to unprofiled tasks without re-measurement. Equally significant is what the validation reveals about failure modes. The accuracy-greedy configuration exceeds the deadline on every cycle, driven not by resource exhaustion but by the model's inconsistent output behavior. On 24\% of input frames, the model produces unexpectedly verbose outputs that inflate VLM latency well beyond the profiled mean. Standard benchmarks with curated inputs do not expose this behavior, but it appears immediately under the sustained, diverse-input conditions typical of CPS deployment.

\textit{Limitations and Extensibility.}
The composability validation in Section~\ref{sec:eval-composability} is conducted under the HuggingFace Transformers inference backend on a single device. Concurrent execution, dynamic batching, and validation under optimized runtimes are reserved for future work. Further, the LLM-as-a-judge component is used without calibration against human ratings or a second judge. Variations in judgment strategy are orthogonal to the characterization workflow itself, and alternative judge models, prompting schemes, and safety- or privacy-oriented scoring policies can be integrated through HoliBench's existing evaluation interface contracts, allowing those research directions to build on the framework's multi-level measurement output without modifying the underlying architecture. Finally, HoliBench standardizes measurement semantics across platforms through consistent energy integration, idle subtraction, and temporal alignment. It does not claim absolute equivalence between underlying telemetry interfaces which expose measurements at different scopes (e.g., GPU, SoC, or board level). We note that real world tasks with agents often employ large batch sizes and generally include prefix-caching for agentic workloads—as such, there is ongoing work to provide tools to analyze cache hit rate across concurrencies and different quantization regimes to extend Holibench's work to large scale agentic workloads. An important limitation is that MLX support has not been added, which would improve the M2 Mac's results.

As foundation models move from cloud APIs to on-device deployment, the gap between capability evaluation and deployment feasibility widens. HoliBench addresses this gap with reusable measurement infrastructure, modality-specific profiling, and constraint-aware configuration selection. Our evaluation across multiple models, quantizations, and devices demonstrates that deployment is a hardware-software co-design problem. The best configuration for a given task is the one whose execution profile maps most efficiently onto the target hardware's capabilities and constraints.

We open-source HoliBench to provide the embedded systems and CPS community with standardized instrumentation for foundation model deployment. Future work will extend the framework to support multi-model concurrency as a first-class optimization target and distributed model execution across multiple devices.